\documentclass[a4paper,11pt]{article}
\usepackage{setspace}
\usepackage[left = 2cm, right=2cm, top=3cm, bottom=3cm]{geometry}
\usepackage[pdftex]{graphicx}
\usepackage{cite}
\usepackage{bm}\usepackage{pdfpages}
\usepackage{float}
\usepackage{amsmath,amssymb,latexsym}
\usepackage{color}
\usepackage{xcolor}
\usepackage[T1]{fontenc}
\usepackage{wasysym}
\usepackage{siunitx}
\usepackage{setspace}
\usepackage{relsize}
\usepackage{tikz}
\usepackage{lineno, blindtext}
\usepackage{authblk}
\usepackage[symbol]{footmisc}
\usepackage{lineno}
\usepackage[pdfencoding=auto, psdextra]{hyperref}

\usepackage{mismath}
\usepackage{upgreek}
\begin{document}

	\renewcommand{\figurename}{Fig.}
	
	\title{\color{blue}\textbf{Stiffness-dependent Dielectric Relaxation in Thermoreversible Microgels: Effects of Temperature and Strain}}
	\author[1]{Sayantan Chanda}
	\affil[1]{\textit{Soft Condensed Matter Group, Raman Research Institute, C. V. Raman Avenue, Sadashivanagar, Bangalore 560 080, INDIA}}
	\author[1]{Sonali Vasant Kawale}
	\author[1,*]{Ranjini Bandyopadhyay}
	
	\footnotetext[1]{Corresponding Author: Ranjini Bandyopadhyay; Email: ranjini@rri.res.in}
	\maketitle
	\begin{abstract}
We investigate how particle stiffness governs length-scale-dependent relaxation dynamics in dense suspensions of thermoresponsive PNIPAM microgels subjected to temperature variations and large amplitude oscillatory shear. By combining dielectric spectroscopy with rheometric measurements, we directly correlate microscopic polarization fluctuation dynamics with macroscopic mechanical relaxation. Highly crosslinked, stiffer microgels exhibit longer dielectric relaxation times across a broad temperature range and strain-induced slowing down, whereas softer, open-network microgels exhibit nearly strain-independent dielectric dynamics. Below the volume phase transition temperature (VPTT), macroscopic bulk stress relaxation experiments reveal that suspensions of the softest particles relax most rapidly. For the stiffer microgel suspensions, however, the dynamics decouple: bulk stress relaxation speeds up with increasing stiffness even as dielectric relaxation slows down. Above the VPTT, softer particles undergo greater volume collapse, leading to very rapid bulk stress relaxation. Besides exhibiting slower stress decay, stiffer microgel suspensions also undergo shear-induced structural regeneration during the later stages of deformation. These findings demonstrate that particle stiffness and internal architecture govern relaxation pathways across microscopic and macroscopic length scales, thereby providing a framework for designing adaptive soft metamaterials, stress-dissipative coatings, and self-healing robotic materials.



 \end{abstract}

\section{Introduction:}
Soft materials such as polymer solutions, colloidal suspensions, emulsions, and gels exhibit a rich spectrum of non-equilibrium behaviors when subjected to external perturbations such as temperature changes, electric or magnetic fields, and mechanical deformation~\cite{pusey1986phase, liu1998jamming, mattsson2009soft,romeo2010temperature, pham2006yielding, pham2008yielding, bandyopadhyay2010stress}. Microscopic structural rearrangements driven by these external stimuli dictate macroscopic properties and produce emergent phenomena like phase transitions, flow-induced structural transformations, yielding, shear thinning, and shear thickening~\cite{cai2011mechanics, schild1992poly, pham2008yielding, ahuja2020two, ryder2006shear, hu1994shear, badiger2003shear, stieger2003shear}.
Consequently, understanding how external perturbations alter particle-scale dynamics, and how those dynamics influence bulk rheology, is essential for linking microscopic structural evolution to macroscopic material response ~\cite{pham2008yielding, misra2023dichotomous}.

Owing to their thermoresponsive nature and tunable mechanical properties, soft colloidal poly(N-isopropylacrylamide) (PNIPAM) microgels are a widely-used model system in the study of deformable and compressible materials\cite{snowden1996colloidal, romeo2010temperature, mattsson2009soft, vlassopoulos2014tunable, franco2021glass, franco2025soft}.
As polymer-solvent interactions become increasingly unfavorable with temperature\cite{heskins1968solution, schild1992poly, romeo2010temperature, franco2025soft}, PNIPAM particles in aqueous media undergo a reversible swelling-deswelling transition at a characteristic volume phase transition temperature (VPTT). 
A notable advantage of PNIPAM microgels is that their stiffness can be systematically tuned by incorporating a polar crosslinking agent, such as N, N$^\prime$-methylenebisacrylamide (MBA), during synthesis \cite{ks2025revisiting, misra2024effect}. Higher crosslinker concentration produces stiffer particles\cite{misra2024effect} with significant alterations in
internal architecture~\cite{varga2001effect}. While softer PNIPAM particles have a smaller, lightly-crosslinked core and an extended peripheral corona, stiffer particles are characterized by larger, highly-crosslinked cores and thinner coronas\cite{vlassopoulos2014tunable, su2014dielectric}.
Depending on particle concentration, stiffness, and temperature, dense PNIPAM suspensions can exist in a rich variety of viscoelastic states, including fluid-like phases, glass{es}, attractive gels, and jammed regimes\cite{romeo2010temperature, wang2014revisit, ghosh2019linear, franco2021glass, misra2024effect}. Strong correlations between internal particle deformability and collective suspension dynamics make PNIPAM microgels an excellent platform for probing how microscopic structure governs macroscopic rheology \cite{franco2021glass, scotti2020flow, misra2023dichotomous}.

Dielectric spectroscopy is a powerful probe of molecular and dipolar dynamics~\cite{nicolai1998dynamics, chrissopoulou2015effects, fullbrandt2012probing}. By monitoring polarization fluctuations in response to an applied electric field, these measurements directly reveal local relaxation processes that complement macroscopic insights from conventional rheology. When combined with shear deformation, dielectric spectroscopy allows the simultaneous investigation of mechanical and dielectric responses under nonequilibrium conditions, directly linking microscopic relaxation dynamics to macroscopic flow behavior ~\cite{capaccioli2007applications, watanabe2003rheo, peng2005stress, misra2023dichotomous}. 
This simultaneous access to microscopic relaxation dynamics and macroscopic mechanical behavior is particularly valuable for dense suspensions and jammed systems, where collective particle interactions produce complex viscoelastic behavior ~\cite{steinhauser2016carbon, misra2023dichotomous}.  

The temperature-dependent structural evolution of PNIPAM microgels has been extensively investigated using dielectric spectroscopy, which is highly sensitive to polymer and dipolar relaxation processes associated with particle swelling and collapse \cite{su2014dielectric, su2016influence, yang2017relaxations}. 
These studies demonstrate that relaxation behavior depends strongly on the internal crosslinker distribution, clearly distinguishing core-corona structures from more homogeneous microgel morphologies \cite{su2014dielectric}.
Simultaneously, bulk rheological studies have revealed strong correlations between particle stiffness, concentration, and the emergence of viscoelasticity and yielding in dense PNIPAM suspensions\cite{pham2006yielding, pham2008yielding, behera2017effects, franco2021glass, franco2025soft, chanda2026investigating}. Recent work from our group demonstrates that particle stiffness controls both the microstructural organization and phase behavior of dense PNIPAM suspensions \cite{misra2024effect}. Another study reported that increasing the softness of PS–PNIPAM microgels by increasing the shell-to-core ratio shifts the yielding transition to larger strains, highlighting the crucial role of shell deformability in the mechanical response of soft glasses. \cite{zhou2014yielding}. 
However, despite several independent dielectric and bulk rheological studies, the relationship between microgel particle stiffness, dielectric relaxation of dense suspensions under applied deformation, and its correlation  with macroscopic stress relaxation dynamics remains largely unexplored.

We employ {rheo-}dielectric spectroscopy to investigate temperature- and strain-dependent dielectric relaxations in dense suspensions of PNIPAM microgels of different particle stiffnesses. Additionally, we measure the macroscopic stress relaxation moduli, which we correlate with the observed dielectric signatures. Dielectric relaxation timescales slow down with {increase in} particle stiffness {and applied external deformation} across the temperature range studied, which we attribute to changes in the core-corona morphology and effective suspension volume fraction. 
Interestingly, suspensions composed of very soft particles with open core-corona architectures display minimal strain dependence.

Below the VPTT, bulk stress-relaxation measurements show that suspensions of the softest particles relax most rapidly, as their high deformability enables them to slip past one another under strain.
However, as particle stiffness increases, bulk relaxation becomes considerably more sluggish before eventually speeding up, even as the dielectric response continues to slow down. 
This contrasting behavior originates from the different length scales probed by our dielectric spectroscopy and rheology experiments.
The very low deformability of the stiffest particles promotes disruption of the system-spanning network under applied strain, resulting in faster macroscopic stress relaxation. Particles of intermediate stiffness, in contrast, have a more lightly crosslinked core and extended corona. They remain deformable, although to a lesser extent than the softest particles which are too compliant to sustain a system-spanning network of inter-particle contacts under strain. Particles of intermediate stiffness can therefore deform to adapt to neighboring particles, forming transient networks that relax via very slow particle rearrangements\cite{scotti2020flow, ikeda2012unified}.
Above the VPTT, particle collapse decreases the effective volume fraction, driving rapid relaxation in soft microgel suspensions. Stiffer suspensions experience less collapse and thus a smaller drop in volume fraction, resulting in a more gradual initial stress decay followed by shear-induced structural reformation. Overall, microgel stiffness emerges as a key parameter governing the dynamics of PNIPAM particles in aqueous suspensions across microscopic and macroscopic length scales, providing design principles for tunable, stimuli-responsive soft materials, and applications 
such as stress-dissipative coatings, flow batteries
~\cite{helal2016simultaneous} and flow capacitors ~\cite{presser2012electrochemical}.

\section{\label{em}Materials and methods}
\subsection{{PNIPAM s}ynthesis and sample preparation:}
N-isopropylacrylamide (NIPAM, 99\%), N, N$^{\prime}$-methylenebisacrylamide (MBA, 99.5\%), sodium dodecyl sulfate (SDS), and potassium persulfate (KPS) (99.9\%) were procured from Sigma-Aldrich and used without further purification. The synthesis procedure followed in this study was the same as reported in previous work\cite{mcphee1993poly,ks2025revisiting,misra2024effect}.
Briefly, 7.0 gm of the monomer, NIPAM, and 0.03 gm of SDS were dissolved in 470 mL of Milli-Q water (Millipore Corp.) in a three-necked round-bottom flask. 
Four separate batches of PNIPAM particles with different stiffnesses were synthesized by adding predetermined amounts of crosslinker (MBA): 0.035, 0.070, 0.105, and 0.140 gm corresponding to crosslinker-monomer concentration ratios of 5\%, 10\%, 15\%, and 20\%, respectively. Particle stiffness increases with increase in crosslinker-monomer concentration ratio\cite{misra2024effect}.
The solutions were stirred at 600 rpm and purged with nitrogen gas for 30 min to establish an inert atmosphere prior to polymerization.
The reaction temperature was subsequently increased to 70${^\circ}$C, and 0.28 gm of KPS{,} dissolved in 30 mL of Milli-Q water{,} was added to initiate polymerization. Th{e quantity of initiator used} was the same {for} synthesis of particles of all stiffnesses. The reaction was allowed to proceed for 4 h under continuous stirring. Upon completion, the suspensions were cooled to room temperature. 
Purification of the synthesized particles was performed by four successive centrifugation cycles at 20,000 rpm for 60 min each. After every centrifugation step, the supernatant was discarded and replaced with fresh Milli-Q water to remove residual SDS, unreacted monomers, and other impurities. Following the final centrifugation cycle, the purified particles were dried completely and subsequently ground into a fine powder.
Dense aqueous PNIPAM suspensions were prepared by dispersing predetermined amounts of the dry PNIPAM powder in Milli-Q water. The suspensions were gently stirred for 48–72 h to ensure complete homogenization and then stored at 4${^\circ}$C for future use ~\cite{misra2020influence, misra2024effect}.


\subsection{Field emission scanning electron microscopy (FESEM):}
{
FESEM was performed to observe changes in the core-corona morphology of dry PNIPAM particles as a function of particle stiffness. A field-emission scanning electron microscope 
(Ultra Plus FESEM-4098; Carl Zeiss) 
with an electron beam energy of 5 KeV {was used}. 10$\mu$l drops containing aqueous PNIPAM suspensions of concentration
$\sim$ 0.01 wt\% were cast on an indium tin oxide (ITO){-}coated glass slide and allowed to dry overnight inside a closed Petri dish under ambient conditions. 
The ITO substrate was then loaded on the FESEM stage for imaging. 
A 3 nm platinum coating was applied to the sample surfaces to ensure good image contrast. Back-scattered secondary electrons were used to reconstruct the surface images of the samples{. Representative images of dry PNIPAM particles of different stiffnesses are} displayed in Fig.~S1 of the supplementary material. A representative analysis of the core-shell particle structure is presented in Fig.~S2 and will be discussed later.
}

\subsection{Dynamic light scattering (DLS):}
{

The mean hydrodynamic diameters, $<$$\mathrm{d_{h}}$$>$, of non-interacting PNIPAM particles dispersed in dilute aqueous suspensions were determined as a function of temperature using dynamic light scattering (DLS) measurements. The sample temperature was regulated between 10$^\circ$C and 50$^\circ$C.
Intensity autocorrelation functions, C({$t$}), over delay times {$t$} ranging from 0.1 $\mu$s to 500 $ m$s were obtained and subsequently analyzed to determine $<$$\mathrm{d_{h}}$$>$. A detailed description of the analysis procedure of $<$$\mathrm{d_{h}}$$>$ is provided in {s}ection ST2 of the {s}upplementary {m}aterial. Additional details regarding the experimental setup can be found elsewhere \cite{mymanual,saha2014investigation, behera2017effects}. The temperature-dependent mean hydrodynamic diameters, $<$$\mathrm{d_{h}}$$>$, of particles of different stiffnesses are shown in Fig. S4 of the supplementary material. 

Since PNIPAM particles are compressible and deformable, they can pack beyond the random close packing limit of hard spheres, making the conventional volume fraction, $\phi$, an inadequate measure of particle packing \cite{senff1999rheology}. We therefore {computed} a modified parameter, the effective volume fraction $\phi_{{eff}}$, using a protocol described in {s}ection ST3 of the {s}upplementary {m}aterial. Fig. S6 of the supplementary material shows the temperature dependence of $\phi_{{eff}}$ for suspensions composed of particles {having distinct} stiffness {values}. 
}  

\subsection{Rheo-dielectric and bulk stress relaxation measurements:}
{
Dielectric measurements were performed using a precision impedance analyzer (Wayne Kerr Electronics, 6500B series) integrated into a stress-controlled Anton Paar MCR 702 rheometer~\cite{macosko1994rheology, misra2023dichotomous}. A parallel-plate geometry (PP-25/DI/TI) with a plate diameter of 25 mm and a gap of 0.5 mm was employed. Dense aqueous PNIPAM suspensions were loaded between the two plates, which simultaneously served as a rheometer shearing cell and capacitor plates. 
{D}ielectric measurements were {performed} over a temperature range between 5$^\circ$C and 45$^\circ$C in the absence of any externally applied shear. At each temperature, an alternating voltage of amplitude 750 mV was applied over a frequency range of 20 Hz to 5 MHz, and the capacitance (C) was measured using the impedance analyzer. The measured capacitance values were then used to calculate the real part of the relative dielectric permittivity ($\mathrm{\upepsilon}^\prime_\mathrm{r}$): 
$\mathrm{\upepsilon}^\prime_\mathrm{r}=\frac{\mathrm{Cs}}{\mathrm{A}\mathrm{\upepsilon}_\mathrm{0}}$,
where s is the plate separation, A is the plate area, and $\mathrm{\upepsilon}_\mathrm{0}$ is the permittivity of free space.
Subsequently, rheo-dielectric measurements were performed below and above the suspension VPTT at 20$^\circ$C and 45$^\circ$C {respectively}. In these experiments, oscillatory shear strains of peak-to-peak amplitude ranging from 0\% to 2000\% were applied at a fixed angular frequency of 10 rad/s{, and} $\mathrm{\upepsilon}^\prime_\mathrm{r}$ was simultaneously measured {between} 20 Hz {and} 5 MHz. This approach allowed us to investigate the influence of shear-induced microstructural changes on the dielectric response of {dense} PNIPAM suspensions. {Stress relaxation moduli were measured in }large amplitude oscillatory shear (LAOS) experiments {and correlated with the} dielectric responses {of the samples}. By correlating bulk stress relaxation with rheo-dielectric observations, we systematically studied how particle stiffness influences the coupling between microstructural dynamics and macroscopic mechanical properties. All temperature and strain-dependent rheo-dielectric and bulk stress relaxation measurements were performed with dense suspensions of particles of {different} stiffnesses prepared at $\phi_{{eff}}$ = 1.7 at 20$^\circ$C. We checked the consistency of all our results by performing experiments with samples prepared at a lower effective volume fraction, $\phi_{{eff}}$ = 1.4 at 20$^\circ$C.

}

\section{\label{r&d}Results and Discussion}
\subsection{Effect of crosslinker concentration on particle morphology and thermoresponsivity over a broad temperature range} 

\begin{figure}[!t]
     \centering
     \includegraphics[width=0.75\linewidth]{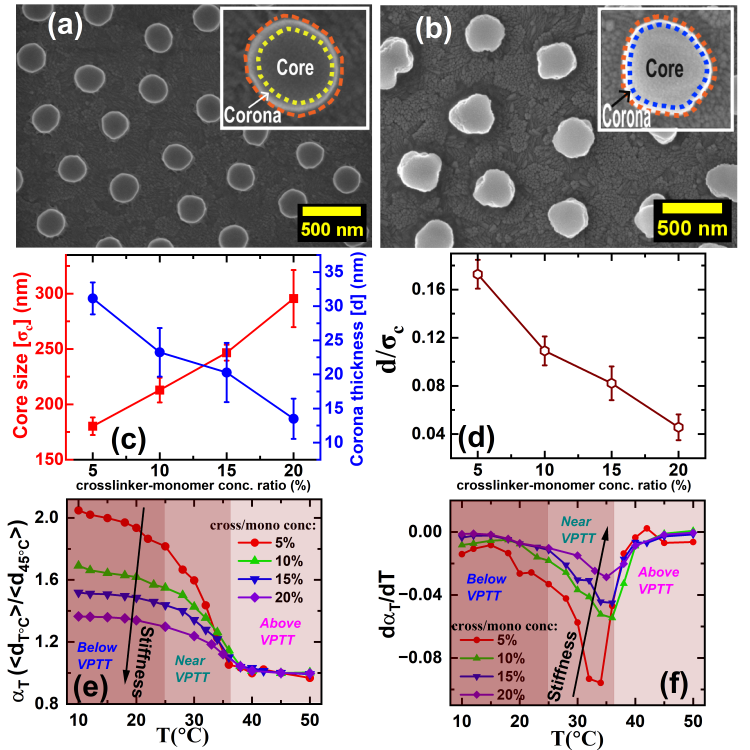}
     \caption{(a) FESEM image of the softest PNIPAM particles synthesized using 5\% crosslinker-monomer concentration ratio. The inset shows a magnified image of a single representative particle, with the region enclosed by the yellow dashed circle corresponding to the central core and the region between the yellow and red dashed circles representing the peripheral corona.  (b) FESEM image of the stiffest PNIPAM particles synthesized using 20\% crosslinker-monomer concentration ratio. The inset shows a magnified image of a representative particle, with the region enclosed by the violet dashed circle corresponding to the core, and the region between the violet and red dashed circles representing the relatively thinner corona. (c) Core size, $\upsigma_\mathrm{c}$, and corona thickness, d, of particles of different stiffnesses estimated from FESEM images. (d) Ratio of corona thickness to core size, d/$\upsigma_\mathrm{c}$, with increasing {particle} stiffness. (e) Temperature-dependent swelling ratio ($\mathrm{\alpha}_\mathrm{T}$=$<$$\mathrm{d}_{\mathrm{T}^\circ \mathrm{C}}$$>$/$<$$\mathrm{d}_{45^\circ \mathrm{C}}$$>$) and (f) d${\mathrm{\alpha}_\mathrm{T}}$/d$\mathrm{T}$ of PNIPAM particles of different stiffnesses ({labeled as} cross/mono conc in figure legends) {suspended} in {an} aqueous medium. Black arrows in (e) and (f) point toward the direction of increasing stiffness.}
     \label{1}
 \end{figure}

Figure~\ref{1}(a) {displays} an FESEM image of the softest microgels synthesized with 5\% crosslinker-monomer concentration ratio. 
The magnified image of an individual particle in the inset clearly reveals a core–corona morphology, with the core highlighted by a yellow dashed circle and the corona spanning the region between the yellow and red dashed circles. Figure~\ref{1}(b) shows the stiffest microgels synthesized with 20\% crosslinker-monomer concentration ratio, where the magnified image in the inset reveals a larger core (highlighted by the violet dashed circle), and a significantly thinner corona between the violet and red dashed circles. 
Additional FESEM images are displayed in Fig.~S1 of the supplementary material. The protocol to quantify the core-corona morphologies is illustrated in Figs.~S2(a-c) of the supplementary material. Figure~\ref{1}(c) shows that the core size, $\upsigma_\mathrm{c}$, increases, while the corona thickness, d, decreases systematically with {increasing crosslinker-monomer concentration ratio, and therefore} increasing particle stiffness. Consequently, the ratio d/$\upsigma_\mathrm{c}$ decreases monotonically with increasing stiffness, as shown in Fig.~\ref{1}(d).  
Increasing crosslinker content therefore transforms the internal particle architecture from a heterogeneous core–corona morphology to a more homogeneous and compact structure, consistent with previous reports\cite{ks2025revisiting, misra2024effect}.

The effect of particle stiffness on the thermoresponsive behavior of PNIPAM microgels was next investigated by measuring the temperature-dependent mean hydrodynamic diameter, $<$$\mathrm{d_{h}}$$>$, using dynamic light scattering (DLS). Figure S4 of the supplementary material shows $<$$\mathrm{d_{h}}$$>$ as a function of temperature for microgels synthesized using {several different} crosslinker-monomer concentration ratios. We observe that particles of all stiffnesses exhibit the characteristic volume phase transition, which is marked by a sharp decrease in size near the VPTT. Consistent with previous studies\cite{ks2025revisiting, misra2024effect}, the VPTT increases slightly with increasing particle stiffness in the range between 32 - 35$^\circ$C.
Figure~\ref{1}(e) displays the swelling ratio, ${\mathrm{\alpha}_\mathrm{T}}$, defined as the ratio of the mean hydrodynamic diameter{s} at a given temperature T, $<$$\mathrm{d}_{\mathrm{T}^{\circ}\mathrm{C}}$$>$, to that in the fully shrunken state at 45$^{\circ}$C, $<$$\mathrm{d}_{45^{\circ}\mathrm{C}}$$>$\cite{misra2024effect}. 
The temperature-sensitivity of ${\mathrm{\alpha}_\mathrm{T}}$ is weaker for particles synthesized with higher crosslinker content, indicating that increased crosslinking yields stiffer particles with denser, more compact internal polymer networks.\cite{ks2025revisiting, misra2024effect}. A larger variation in $\alpha_\mathrm{T}$ with temperature for less crosslinked softer particles reflects a more open polymer network, which undergoes a greater volume change upon thermal collapse.

A previous study ~\cite{daly2000temperature} demonstrated that the central core of PNIPAM particles undergoes an initial gradual collapse as the temperature is increased below the VPTT. A more pronounced reduction in particle size near the VPTT is attributed to a rapid collapse of the peripheral corona. At temperatures above the VPTT, the core collapse{s further} as the particle approaches its fully shrunken state.
We see from Fig.~\ref{1}(f), which displays the temperature-derivative of ${\mathrm{\alpha}_\mathrm{T}}$ as a function of particle stiffness, that softer particles exhibit a pronounced change in d${\mathrm{\alpha}_\mathrm{T}}$/d$\mathrm{T}$ even well below the VPTT, indicating a weakly crosslinked core that begins shrinking at lower temperatures.
In contrast, the response of stiffer particles below the VPTT is much weaker, consistent with a denser core that resists thermal collapse. Near the VPTT, the broader and {larger} variation of d${\mathrm{\alpha}_\mathrm{T}}$/d$\mathrm{T}$ in softer particles indicates {a gradual collapse of their} spatially extended coronas over a wider temperature range. The weaker and narrower response of stiffer particles under the same conditions {highlights} the {rapid} collapse of their thinner coronas over a smaller temperature interval. Together, these observations demonstrate that soft particles, with their dilute cores and extended coronas, undergo pronounced deswelling with increasing temperature, whereas stiffer particles {with} denser, more homogeneous morphologies, deswell much less. 

\subsection{Effect of particle stiffness on temperature-dependent dielectric {response}}

\begin{figure*}[th]
     \centering
     \includegraphics[width=0.4\linewidth]{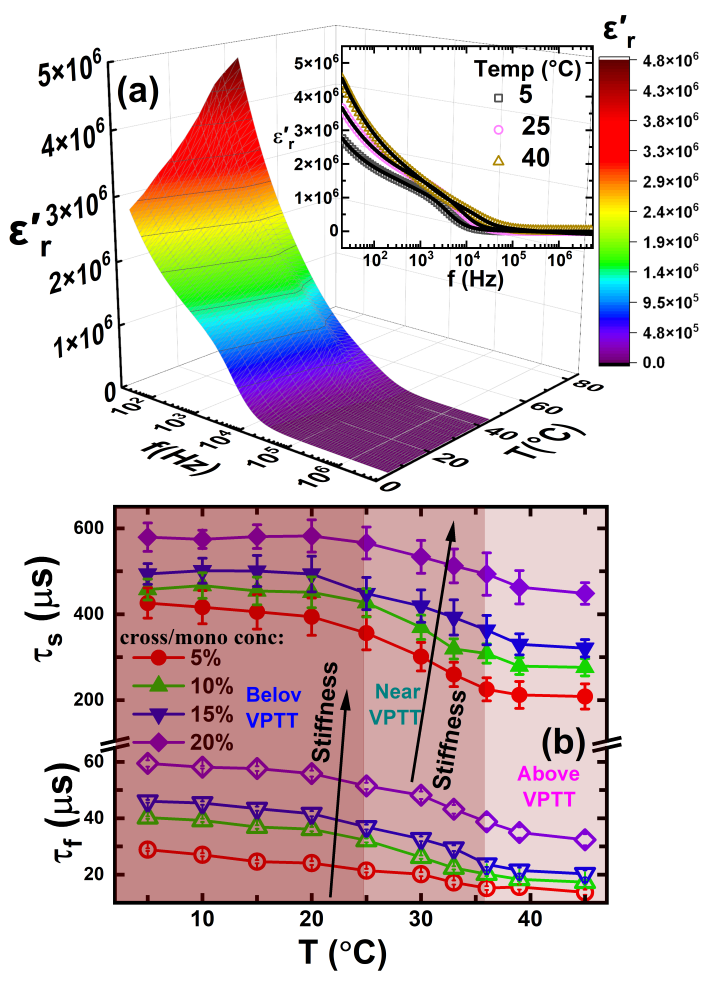}
     \caption{(a) Three-dimensional representation of {the} real part of the relative dielectric permittivity, $\mathrm{\upepsilon}^\prime_\mathrm{r}$, of dense aqueous suspension of PNIPAM particles ($\phi_{eff}$ = 1.7 prepared at 20${^\circ}$C) with 15\% crosslinker-monomer concentration ratio as a function of {applied frequency and} temperature. The inset shows $\mathrm{\upepsilon}^\prime_\mathrm{r}$ vs frequency at three representative temperatures. The black lines are fits to the real part of the Cole-Davidson model given in Eqn.~\ref{eqn}. (b) Temperature-dependent slow relaxation timescale{s}, $\uptau_\mathrm{s}$ (bold symbols),  and fast relaxation timescale{s},  $\uptau_\mathrm{f}$ (hollow symbols), of particles {of} different stiffnesses in the aqueous suspensions. 
     Black arrows in (b) point in the direction of increasing particle stiffness.
     }
     \label{2}
 \end{figure*}

The influence of particle stiffness on the temperature-dependent microscopic dynamics of dense suspensions of thermoresponsive PNIPAM microgels ($\phi_{eff}$ = 1.7 at 20${^\circ}$C) was investigated using dielectric spectroscopy 
over a broad temperature range. 
Figure~\ref{2}(a) shows a three-dimensional representation of the real part of the {relative} dielectric permittivity ($\mathrm{\upepsilon}_\mathrm{r}'$) for a dense aqueous suspension of PNIPAM particles, prepared with 15\% crosslinker-monomer concentration ratio, as a function of applied frequency and temperature under zero strain conditions. 
Consistent with previous studies ~\cite{su2014dielectric, misra2023dichotomous}, $\mathrm{\upepsilon}_\mathrm{r}'$ exhibits two distinct relaxation processes in the low and high frequency regimes throughout the temperature range explored.
The inset of  Fig.~\ref{2}(a) {displays representative data for the} frequency response of $\mathrm{\upepsilon}^\prime_\mathrm{r}$ at three different temperatures, {and the corresponding fits (black lines) to} the real part of the complex Cole–Davidson dielectric model~\cite{su2014dielectric, yang2017relaxations}{:}
\begin{equation}
\mathrm{\upepsilon}^{*}
=
\mathrm{\upepsilon}_\mathrm{{h}}
+
\frac{\mathrm{\upepsilon}_\mathrm{{l}}-\mathrm{\upepsilon}_\mathrm{{m}}}
{1+\left(j\omega \uptau_{s}\right)^\mathrm{{\beta_{s}}}}
+
\frac{\mathrm{\upepsilon}_\mathrm{{m}}-\mathrm{\upepsilon}_\mathrm{{h}}}
{1+\left(j\omega \uptau_\mathrm{{f}}\right)^\mathrm{{\beta_{f}}}}
+
A\omega^{\mathrm{-m}}
  \label{eqn}
\end{equation}
{Here}, $\mathrm{\upepsilon}_\mathrm{l}$, $\mathrm{\upepsilon}_\mathrm{m}$, and $\mathrm{\upepsilon}_\mathrm{h}$ represent the dielectric permittivities in the low, intermediate, and high-frequency limits, respectively.  $\uptau_\mathrm{s}$ and $\uptau_\mathrm{f}$ denote the characteristic relaxation times of the slow and fast processes, while $\beta_\mathrm{s}$ and $\beta_\mathrm{f}$ {are} the corresponding stretching exponents. 
 The last term in Eqn.~{1}, $A\omega^{\mathrm{-m}}${,} accounts for electrode polarization (EP) at lower frequencies due to the accumulation of free charges at the electrode interface~\cite{yang2017relaxations, misra2023dichotomous}. 
The slow relaxation process, characterized by the timescale $\uptau_\mathrm{s}$ and extracted from the fit of the data to Eqn.~(1), is attributed to segmental motion of the polymer chains throughout the temperature range studied ~\cite{yang2017relaxations}. In contrast, the fast relaxation process, having a characteristic relaxation time $\uptau_\mathrm{f}$, is believed to arise from side-chain relaxation and counterion fluctuations below the VPTT ~\cite{su2014dielectric, yang2017relaxations}, and interfacial polarization due to the separation of counterions (introduced by SDS and KPS) above the VPTT~\cite{yang2017relaxations, su2014dielectric, su2016influence}. The complete dataset and the corresponding fits are shown in Fig. S7 of the supplementary material.

Figure~\ref{2}(b) shows that  $\uptau_\mathrm{s}$ decreases {monotonically} with increasing temperature for all samples. Interestingly, suspensions of stiffer microgels, composed of particles with larger, more crosslinked cores and relatively shorter coronas, consistently exhibit larger  $\uptau_\mathrm{s}$. In contrast, suspensions of softer particles, characterized by open, loosely crosslinked core and larger coronas,  exhibit smaller  $\uptau_\mathrm{s}$ values due to greater segmental mobility.
As seen from Fig. S6, stiffer particles undergo a more gradual reduction in {$\phi_{eff}$ with increase in temperature}. The decrease in  $\uptau_\mathrm{s}$ with temperature is therefore less pronounced in stiffer microgels{. This is} highlighted by the weaker {temperature-dependence of} d$\uptau_s/$dT in Fig.~S8 of the supplementary material{, which is believed to} aris{e} from reduced polymer mobility {and} result{s} in {the observed} slowing down of $\uptau_\mathrm{s}$.

The fast relaxation time, $\uptau_\mathrm{f}$, exhibits a similar stiffness{-}dependence. Below the VPTT, the higher polymer densities of stiffer particles restrict side-chain motion, while their higher surface charge densities~\cite{misra2024effect} hinder counterion fluctuations, resulting in larger $\uptau_\mathrm{f}$. Above the VPTT where $\uptau_\mathrm{f}$ is governed by interfacial polarization, suspensions of stiffer particles exhibit slower dynamics {as} their higher $\phi_{eff}$ (Fig. S6) increases electrical double-layer overlap {and slows down} interfacial polarization relaxation.
We conclude therefore that the dielectric relaxation dynamics reflect the combined effects of particle core–corona architecture and temperature-dependent changes in the effective volume fraction. 

\subsection{Influence of large amplitude oscillatory deformation on dielectric relaxation dynamics below the VPTT:} 

\begin{figure*}[th]
     \centering
     \includegraphics[width=0.4\linewidth]{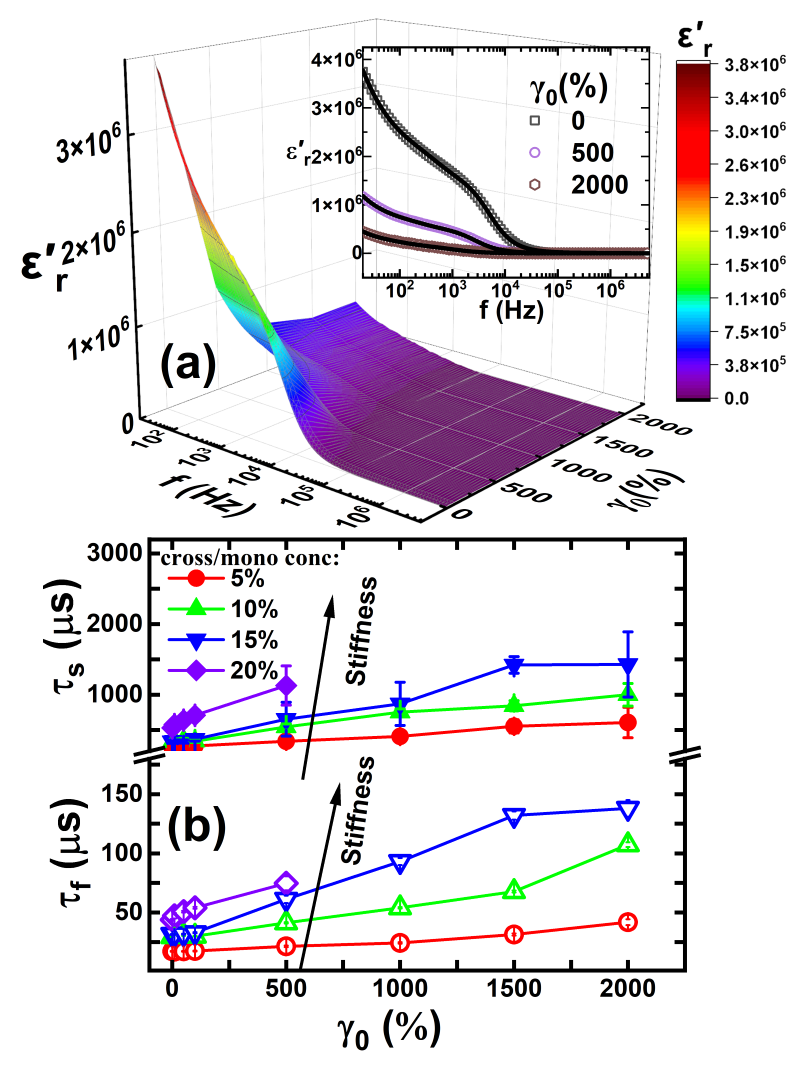}
     \caption{(a) Three-dimensional representation of the real part of the relative dielectric permittivity, $\mathrm{\upepsilon}^\prime_\mathrm{r}$, of dense aqueous suspension of PNIPAM particles ($\phi_{eff}$ = 1.7 prepared at 20${^\circ}$C) with 15\% crosslinker-monomer concentration ratio, as a function of {applied frequency and} strain at 20$^\circ$C. The inset shows $\mathrm{\upepsilon}^\prime_\mathrm{r}$ vs frequency at three representative strain values. The black lines are fits to the real part of the Cole-Davidson model given in Eqn.~\ref{eqn}. (b) Strain-dependent slow relaxation timescales,  $\uptau_\mathrm{s}$ (bold symbols),  and fast relaxation timescales, $\uptau_\mathrm{f}$ (hollow symbols). 
     Black arrows in (b) point in the direction of increasing particle stiffness. }
     \label{3}
 \end{figure*}

To investigate the coupling between microscopic polarization dynamics and large mechanical deformations, rheo-dielectric measurements were performed under large amplitude oscillatory strains. 
Fig.~\ref{3}(a) shows a three-dimensional representation of the frequency- and strain-dependencies of the real part of the dielectric permittivity, $\mathrm{\upepsilon}^\prime_\mathrm{r}$, for a dense aqueous suspension of PNIPAM particles ($\phi_{eff}$ = 1.7 prepared at 20${^\circ}$C) with 15\% crosslinker-monomer concentration ratio below the VPTT (20${^\circ}$C). The inset shows representative dielectric spectra measured at three strain amplitudes and fits of the data (black lines) to the Cole–Davidson model. The entire set of strain-dependent frequency response data below the VPTT and the corresponding fits to Eqn.~1 are shown in Fig.~S9 of the supplementary material.

Figure~\ref{3}(b) shows the strain dependence of  $\uptau_\mathrm{s}$ and $\uptau_\mathrm{f}$ below the VPTT (at 20${^\circ}$C).  
We note that $\uptau_\mathrm{s}$ and $\uptau_\mathrm{f}$ increase monotonically with strain for the suspensions of stiffer particles composed of highly crosslinked microgels, whereas only weak strain dependence is observed for softer particle suspensions even under large deformations. The denser polymer networks in stiffer microgels are more strongly perturbed by shear, restricting segmental and side-chain motion and slowing both relaxation processes. In contrast, the more open core-corona structure of softer microgels favors faster polymeric relaxation such that relaxation timescales remain largely insensitive to strain. 

Figures S10(a) and S10(b) show low and high frequency dielectric strengths, $\Delta\mathrm{\upepsilon}_\mathrm{l}$ and $\Delta\mathrm{\upepsilon}_\mathrm{h}$ respectively, {which} decrease with increasing strain for stiffer particle suspensions but remain nearly unchanged for softer ones. The strong reduction for stiffer particles suggests that {large} deformation{s} suppress polarization fluctuations, supporting our conclusions from relaxation time {data (Fig.~\ref{3}(b))}. {In our experimental arrangement, the shear flow is orthogonal to the electric field. As strain increases, strong shear aligns the dipoles parallel to the flow direction~\cite{misra2023dichotomous, capaccioli2007applications}, thereby decreasing the population of dipoles contributing to polarization and lowering the dielectric strength}. For suspensions of softer particles, nearly constant dielectric strengths indicate that the more deformable and open internal particle structures relax faster than the deformation rate, making them more adaptive than stiffer ones.

\subsection{ Influence of large amplitude oscillatory deformation on dielectric relaxation dynamics above the VPTT:} 

\begin{figure}[th]
     \centering
     \includegraphics[width=0.4\linewidth]{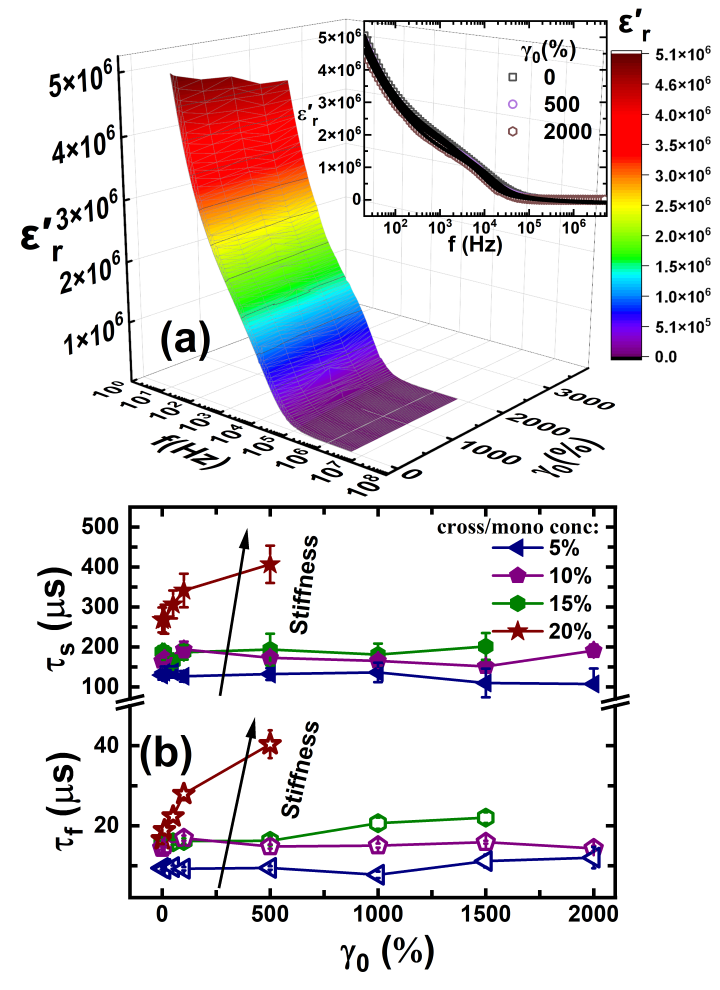}
     \caption{(a) Three-dimensional representation of the real part of the relative dielectric permittivity, $\mathrm{\upepsilon}^\prime_\mathrm{r}$, of a dense aqueous suspension of PNIPAM particles ($\phi_{eff}$ = 1.7 at 20${^\circ}$C){,} prepared with 15\% crosslinker-monomer concentration ratio{,} as a function of {applied frequency and} strain at 45$^\circ$C. The inset shows $\mathrm{\upepsilon}^\prime_\mathrm{r}$ vs frequency at three representative strain values. The black lines are fits to the real part of the Cole-Davidson model given in Eqn.~\ref{eqn}. (b) Strain-dependent slow relaxation timescales,  $\uptau_\mathrm{s}$ (bold symbols),  and fast relaxation timescales, $\uptau_\mathrm{f}$ (hollow symbols). 
     Black arrows in (b) point in the direction of increasing particle stiffness.}
     \label{4}
 \end{figure}

Figure~\ref{4}(a) {presents a three-dimensional plot of the real part of the relative dielectric permittivity, $\mathrm{\upepsilon}^\prime_\mathrm{r}$, versus frequency and strain for a dense aqueous PNIPAM particle suspension} ($\phi_{eff}$ = 1.7 prepared at 20${^\circ}$C) with 15\% crosslinker-monomer concentration ratio above the VPTT (45${^\circ}$C). 
The inset shows representative dielectric spectra measured at three strain amplitudes and {the} fits (black lines) to the Cole–Davidson model (Eqn.~\ref{eqn}).  The complete set of strain-dependent frequency response data above the VPTT and the corresponding fits to Eqn.~\ref{eqn} for all the samples are shown in Fig. S11 of the supplementary material. 

Above the VPTT, relaxation timescales and dielectric strengths are observed to be strain-dependent only in the suspensions composed of the stiffest particles, as displayed in Figs.~\ref{4}(b) and S12 of the supplementary material, respectively.
{Due to} the particle preparation protocol used here, changes in $\phi_{eff}$ are negligible well below the VPTT but depend strongly on particle stiffness above the VPTT (Fig. S6 of the supplementary material). 
{T}he substantially larger $\phi_{eff}$ of suspensions composed of the stiffest particles {severely hinders} the segmental motion of the underlying polymeric structure, leading to the observed strain-dependent increase in  $\uptau_\mathrm{s}$. 
{In contrast, lower $\phi _{eff}$ values in softer particle suspensions reduce particle crowding and prevent self-assembled aggregation, rendering their dielectric responses insensitive to applied deformations.}

Above the VPTT, the fast relaxation timescale, $\uptau_\mathrm{f}$, is dominated primarily by interfacial polarization processes~\cite{su2014dielectric, misra2024effect}.
The higher $\phi_{eff}$ of suspensions constituted by stiffer particles increases the likelihood of overlap between the electrical double layers of neighboring particles. 
Application of strain disrupts the overlapping configurations of the neighboring double layers, leading to a pronounced increase in $\uptau_\mathrm{f}$ only for the stiffest particle {suspensions}. In contrast, suspensions of softer particles have substantially lower $\phi_{eff}$ and are composed of loosely packed structures with minimal double-layer overlap. This renders their dynamics insensitive to applied strain. The dielectric strengths also exhibit behavior consistent with the relaxation timescales. For suspensions of the stiffest particles, both $\Delta\mathrm{\upepsilon}_\mathrm{l}$ and $\Delta\mathrm{\upepsilon}_\mathrm{h}$ decrease with increasing strain amplitude, while negligible variation is observed for softer microgels (Figs. S12(a) and (b) of the supplementary material).
The reproducibility of {all the observed rheodielectric} trends was further confirmed at a lower effective volume fraction, $\phi_{eff}$ = 1.4 prepared at 20${^\circ}$C, across a broad range of applied shears and at temperatures both below and above the VPTT. Th{is data, displayed} in Figs. S13(a-i) of the supplementary material{, dislays significantly shorter} relaxation timescales due to weaker interparticle crowding. 

\subsection{Identifying correlation{s} between dielectric relaxation dynamics and bulk stress relaxation:}

\begin{figure}[!t]
     \centering
     \includegraphics[width=0.4\linewidth]{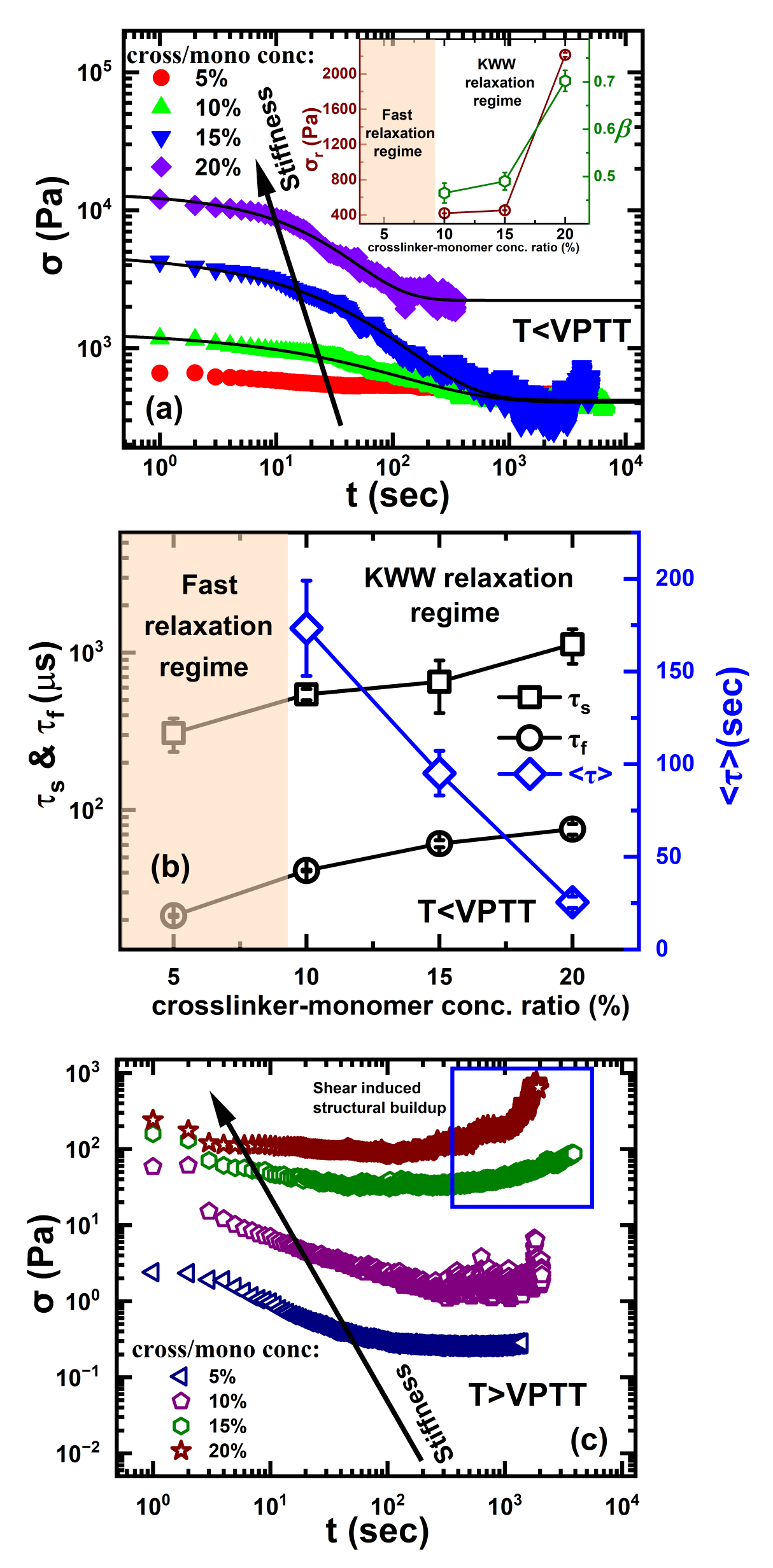}
     \caption{(a) Relaxation of bulk stress with time for dense suspensions ($\phi_{eff}$ = 1.7 prepared at 20${^\circ}$C) {constituted by} PNIPAM particle{s of different} stiffnesses, subjected to a large oscillatory strain of peak-to-peak amplitude, $\gamma_0$ = 500$\%$ below the VPTT (20${^\circ}$C). The inset shows the variation of residual stress, $\upsigma_{\mathrm{r}}$, and stretching exponent, $\beta$, as a function of particle stiffness. (b)  {Slow and fast dielectric relaxation times: }$\uptau_\mathrm{s}$, $\uptau_\mathrm{f}$, and bulk {stress} relaxation time, $<$$\uptau$$>$, as a function of particle stiffness below the VPTT (20$^\circ$C) extracted by fitting the data in (a). (c) Bulk stress relaxation with time for PNIPAM suspensions constituted by particles of {different} stiffnesses, subjected to $\gamma_0$ {=} 500$\%${,} above the VPTT (45${^\circ}$C). {B}lack arrows in (a) and (c) point toward increasing particle stiffness.}
     \label{5}
 \end{figure}

Having established the influence of particle stiffness on dielectric relaxation dynamics, we {next} compare how these microscopic relaxation processes relate to {macroscopic (bulk)} stress relaxation below and above the VPTT. Figure~\ref{5}(a) shows {the} stress relaxation in dense suspensions of PNIPAM particles of {different} stiffness{es, when} subjected to an oscillatory strain of peak-to-peak amplitude $\gamma_{0}$ = 500\% at 20${^\circ}$C. Suspensions composed of the softest particles exhibit the fastest stress relaxation. The low crosslinking density of these particles and their open core-corona structures result in weak network connectivity, enabling rapid particle rearrangement under strain. This claim is supported by {the} cryo-SEM images shown in {Figs.~S14(a-d)}, which clearly reveal markedly thinner network strands in suspensions of {the} softest particles. 
Suspensions with higher particle stiffness{es} exhibit stress relaxation described by stretched exponential function{s:}
$\upsigma(\mathrm{t})=\mathrm{A}\exp[-(\mathrm{t}/\uptau)^\beta ] + \upsigma_{\mathrm{r}}$,
where $\upsigma(\mathrm{t})$ is the time-dependent shear stress, $\uptau$ the characteristic relaxation time, $\beta$ the stretching exponent (lying between 0 and 1), and $\upsigma_{\mathrm{r}}$ the residual stress~\cite{williams1970non}. The mean relaxation timescale is estimated as $<\uptau> =(\frac{\uptau}{\beta})\Gamma(\frac{1}{\beta})$. As shown in the {inset of} Fig.~\ref{5}(a), both $\upsigma_{\mathrm{r}}$ and $\beta$ increase with {particle} stiffness, indicating higher residual stress{es} and narrower distribution{s} of relaxation timescales for suspensions of stiffer particles.

Figure~\ref{5}(b) shows that, {while suspensions of the softest particles exhibit rapid dielectric and stress relaxations, bulk stress relaxation timescales for particles of intermediate and high stiffness show a trend opposite to that observed in the dielectric measurements. This inverse correlation arises from the distinct structural responses probed at the microscopic and macroscopic scales. For particles of intermediate stiffness, the relatively open particle structure undergoes only modest changes in the local dielectric environment under shear, resulting in a weakly varying dielectric response. At the bulk scale, however, the deformable cores and extended coronas readily deform and interpenetrate with those of neighboring particles, increasing interparticle contacts and transient network connectivity. The resulting corona entanglement promotes collective particle rearrangement events, leading to longer bulk relaxation times, $<$$\uptau$$>$, and broader relaxation time distributions} (lower $\beta$ as shown in the inset of Fig.~\ref{5}(a)).

In contrast, {the highly crosslinked cores and short coronas of the stiffer particles have limited deformability and reduced corona interpenetration. U}nder shear{, these particles resist local deformation and instead favor rupture and reorganization of the percolated network into densely packed clusters}~\cite{stieger2003shear, bergman2018new}. These clusters constrain segmental and side-chain motion and enhance local confinement, thereby slowing down the dielectric relaxation times~\cite{misra2023dichotomous}{. At the bulk scale, however, disruption of} the percolated {stress-bearing} networks {facilitates stress release and faster particle rearrangement}, resulting in lower $<$$\uptau$$>$ (Fig.~\ref{5}(b)), and a narrower {relaxation time} distribution, {(high $\beta$,} inset of Fig.~\ref{5}(a)). {Additionally, t}he limited deformability of stiffer particles {results in} higher residual stresses, $\upsigma_{\mathrm{r}}$.
Figures S15 and S16 {display} that increased strain amplitude {reduces} residual stress{, accelerates bulk stress relaxation,} and narrows the timescale distribution. {Larger} applied strain{s} lead to significant rupture of particle networks, leading to faster rearrangement{s and the observed speeding up of the bulk relaxation process}, {in agreement} with previous reports~\cite{misra2023dichotomous}. At lower $\phi_{eff}$, {the} stress decays rapidly for suspensions of softer particles {(Fig.~S17)}, reflecting weak microstructures in these quiescent samples.

Figure~\ref{5}(c) demonstrates that{,} above the VPTT (45${^\circ}$C), soft particle suspensions {undergo rapid} stress {relaxation} due to a substantial reduction in $\phi_{eff}$, while suspensions {of stiffer particles} exhibit slow{er} initial relaxation followed by shear-induced stress growth, indicating delayed structural {regenera}tion. Cryo-FESEM images above the VPTT (Figs.~S14(e–h)) reveal {lower particle} packing density with {decreasing} particle stiffness.
We believe that the enhanced packing densities of stiffer particles in suspension and stronger inter-particle hydrophobic attraction ~\cite{misra2024effect} {promote the reformation of particle networks following shear-induced disruption}. Figures ~\ref{5}(c), S18 {and S19} reveal that lower applied $\gamma_{0}$ and $\phi_{eff}$ both delay {the} onset {of stress growth, and therefore structural regeneration.} {Interestingly}, {signatures of structural regeneration were} never observed below the VPTT, {due to} weaker {inter-particle} hydrophobic attractions ~\cite{misra2024effect}.
Collectively, the rheo-dielectric and stress-relaxation data demonstrate that particle stiffness {govern}s relaxation dynamics {differently} across {microscopic and macroscopic length} scales and temperature regimes.

\section{\label{se:sac} Conclusions:}

This study establishes that particle stiffness, tuned via crosslinker concentration, governs relaxation dynamics across microscopic and macroscopic length scales in dense microgel suspensions.
We observe that {increasing particle stiffness slows dielectric relaxation, both with increase in temperature and under applied strain, while simultaneously accelerating bulk stress relaxation}.
We attribute this dichotomy to the different physical processes and length scales probed by dielectric spectroscopy and rheology. 
{I}ncreas{ing the} crosslinker concentration transforms the particle morphology from a soft core–corona architecture to a denser, more homogeneous structure, {thereby resulting in progressively constrained local polymeric dynamics} with increas{ing} particle stiffness. As the temperature is raised, the highly crosslinked cores and the short corona{s} of stiff particles {lead to a} weaker reduction in $\phi_{eff}${,} and higher {local} polymer density.
{Under shear, the denser polymer networks in stiffer microgels restrict segmental and side-chain motion, slowing down dielectric relaxation below the VPTT.}
Above the VPTT, {the higher incompressibility of these particles results in} higher $\phi_{eff}$ {compared to softer particle suspensions and leads to a strain-induced slowing-down of the dielectric} relaxation processes{. In contrast,} the {larger} reduction in $\phi_{eff}$ for softer particles {renders their} dynamics comparatively insensitive to strain.

At the macroscopic scale, bulk stress relaxation exhibits a distinct non-monotonic behavior below the VPTT, and is slowest in suspensions constituted by intermediate-stiffness particles whose deformability enables collective rearrangements while maintaining a system-spanning network. In contrast, both the softest and stiffest particle suspensions relax rapidly, but \textit{via} distinct mechanisms: the softest microgel suspensions relax quickly due to insufficient inter-particle connectivity, while the limited deformability of the stiffer particles promotes relaxation through network disruption.
Above the VPTT, the large reduction of $\phi_{eff}$ in the soft particle suspensions accelerates stress relaxation, whereas the stiffer particle suspensions exhibit an initial stress decay that is followed by shear-induced stress growth indicative of delayed structural reformation. 

{Our findings reveal {that} microgel stiffness tunes relaxation {in fundamentally} different {ways} across length scales{:} local polymer dynamics govern microscopic relaxation{, while} collective network connectivity govern{s} macroscopic stress relaxation.} {Thus,} a single structural parameter, {the microgel} particle stiffness, can {produce contrasting} relaxation {responses}{  depending on} {the dominant mechanism at the length scale probed}. More broadly, this work highlights the {importance} of simultaneously probing microscopic and macroscopic dynamics to {uncover} length-scale-dependent {responses} in dense, kinetically constrained soft matter, and {provides} a framework for {connecting particle-scale} structure {to} {collective} dynamics in deformable colloidal and glassy systems driven far from equilibrium.




\renewcommand\refname{References}
	
	\bibliographystyle{elsarticle-num}
	\bibliography{main_ref}
    

\section*{Supplementary material}

# ST1 Field emission scanning electron microscopy (FESEM)

## ST1.1 FESEM images of PNIPAM particles of different particle stiffness:

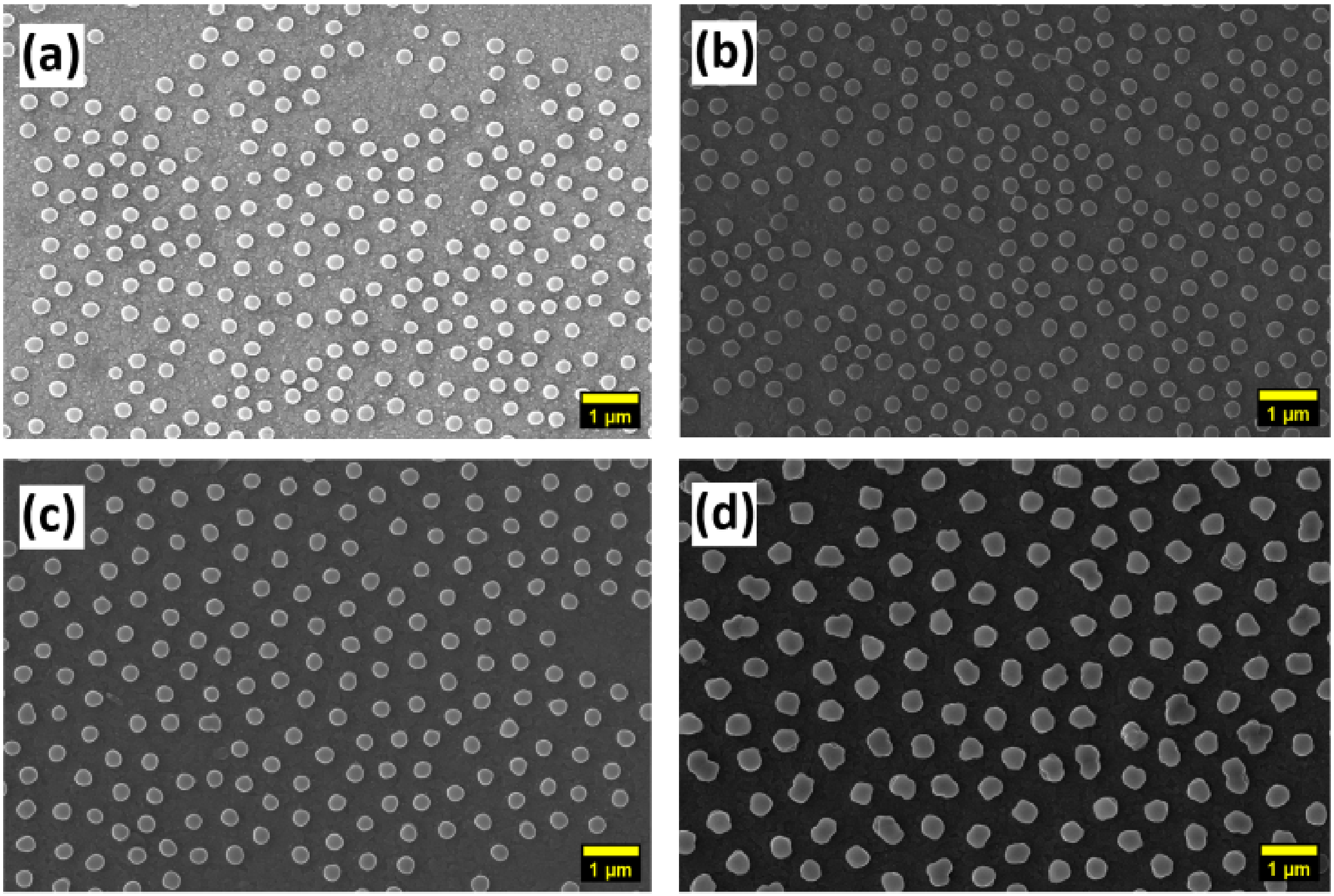


Fig. S1: **FESEM images of dry PNIPAM particles of increasing stiffness: crosslinker-monomer concentration ratios = (a) 5%, (b) 10%, (c) 15% and (d) 20%.**

## ST1.2 Determination of core size, $\sigma_c$, and corona thickness, d, of particles from FESEM images:

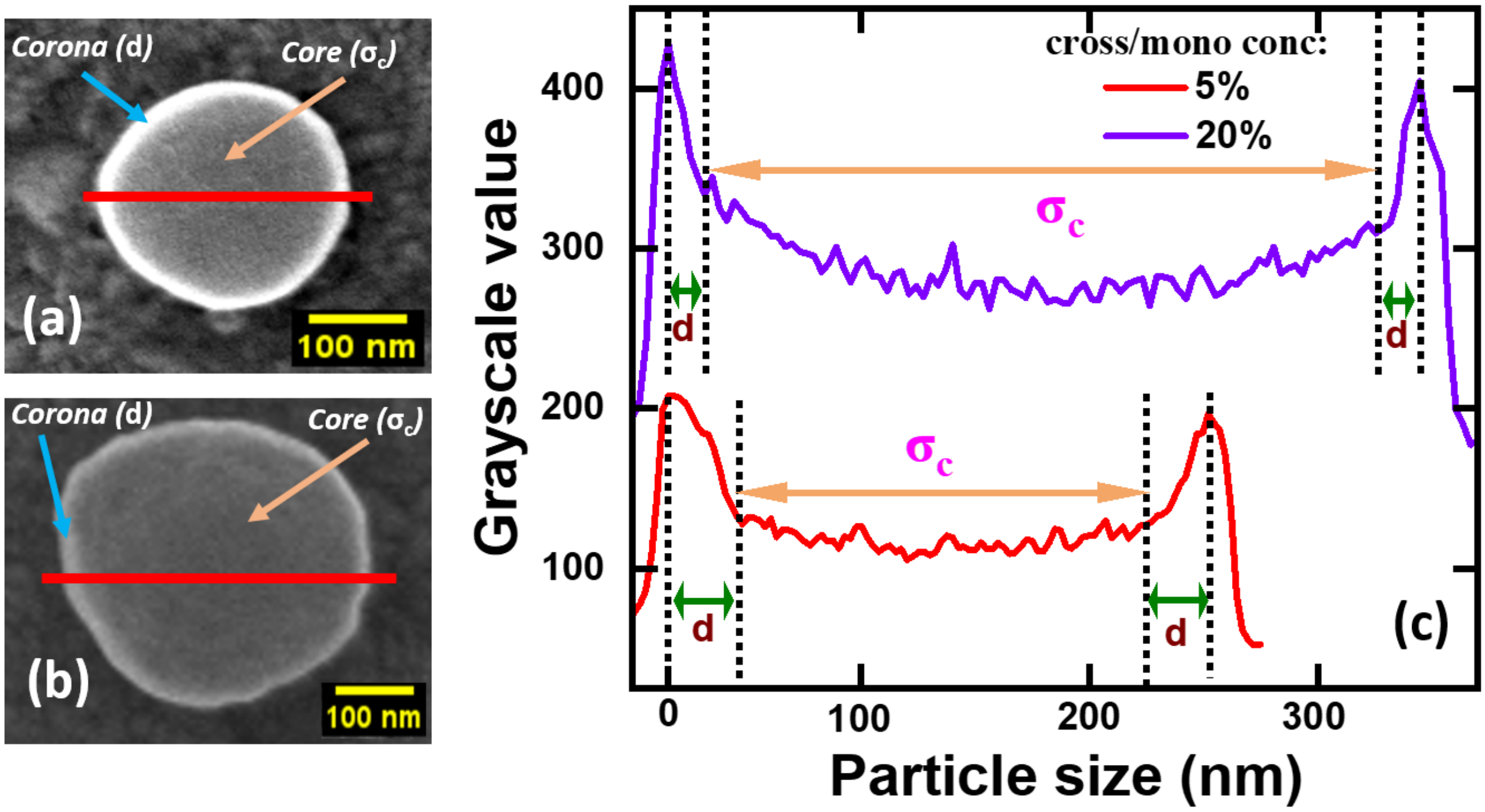


Fig. S2: **Representative FESEM images of PNIPAM particles synthesized with (a) 5% and (b) 20% crosslinker-monomer concentration ratio. A line profile (red line) was drawn across the particle diameter and the grayscale intensity variation was estimated using the *Plot Profile* function in ImageJ. The arrows indicate the central core and peripheral corona. (c) Grayscale intensity profiles obtained from line scans across microgels synthesized with 5% and 20% crosslinker-monomer concentration ratios (labeled as cross/mono conc in figure legend). The positions of the abrupt changes in grayscale intensity were used to determine the apparent core size ($\sigma_c$) and corona thickness. (d) The intensity profile for the particles having 20% crosslinker-monomer concentration ratio has been vertically shifted to avoid overlap with the profile obtained for particles with 5% crosslinker-monomer ratio.**

The core size and corona thickness of the PNIPAM particles were determined from FESEM images using ImageJ (version: 1.49q). For each particle, a line was drawn approximately through its center along the particle diameter as shown by the red line in Figs. S2(a) and S2(b). The grayscale intensity along this line was extracted using the Plot Profile function in ImageJ to obtain the intensity variation across the particle. Figure S2(c) shows the variations in intensity profiles across particles with different crosslinker-monomer concentration ratios. To account for any anisotropy in particle shape, line profiles passing through the particle center were drawn and analyzed along different orientations. To avoid profile overlap and ensure clarity, the grayscale intensity profile corresponding to the particle

synthesized with 20% crosslinker-monomer concentration ratio in Fig. S2(c) has been vertically shifted.

In the FESEM images in Figs. S2(a) and S2(b), the substrate appears as a low-intensity background. This is followed by an abrupt increase in intensity that highlights the bright corona region. Moving further toward the particle center, the intensity decreases abruptly to a relatively uniform lower-intensity region corresponding to the particle core. The grayscale intensity profiles in Fig. S2(c) clearly exhibit these three distinct regions corresponding to the substrate, corona, and core. The positions of these abrupt changes in intensity were taken as the boundaries between the substrate, corona and core, as shown by the black dotted lines in Fig. S2(c). The apparent core size, $\sigma_c$, was defined as the distance between the two inner boundaries, while the corona thickness, d, was determined as the distance between the outer boundary and the corresponding inner boundary on each side. Corona thicknesses and core sizes were estimated by averaging over at least 50 particles for each stiffness value.

## ST2 Determination of the hydrodynamic diameter, <$d_h$>, of PNIPAM particles:

Normalized intensity autocorrelation functions, C($t$)= $\frac{g^{(2)}(q,t)-1}{A}$, were recorded as a function of delay time $t$ using a Brookhaven Instruments digital autocorrelator (BI 9000T). In the expression for C($t$), $A$ is the coherence factor and $g^{(2)}(q,t)$ is defined as $\frac{<I(q,0)I(q,t)>}{<I(q,0)^2>}$, where $I(q,t)$ is the scattered intensity of light at a particular wave vector $q$ and an instant of time $t$. Figure S3 shows C($t$) as a function of $t$ at different temperatures for dilute suspensions of PNIPAM particles synthesized with 15% crosslinker-monomer concentration ratio. For a suspension of polydisperse particles, the temporal decay of C($t$) was modeled as $[\exp(-(\frac{t}{\tau})^{\beta})]^2$. The fitting parameters $\beta$ and $\tau$ were used to calculate the mean relaxation time $<\tau>= (\frac{\tau}{\beta})\Gamma(\frac{1}{\beta})$ [1], where $\Gamma$ is the Euler Gamma function. The inset of Fig. S3 shows the variation of $<\tau>$ with temperature. The mean hydrodynamic diameter, <$d_h$>, of the particles was then estimated using the Stokes-Einstein relationship for spherical particles: $<d_h>= \frac{k_B T<\tau>q^2}{3\pi\eta}$ where $k_B$ is the Boltzmann constant, $T$ is the absolute temperature and $\eta$ is the solvent viscosity. Figure S4 displays the temperature-dependent <$d_h$> of PNIPAM microgels with different particle stiffnesses in aqueous suspension.

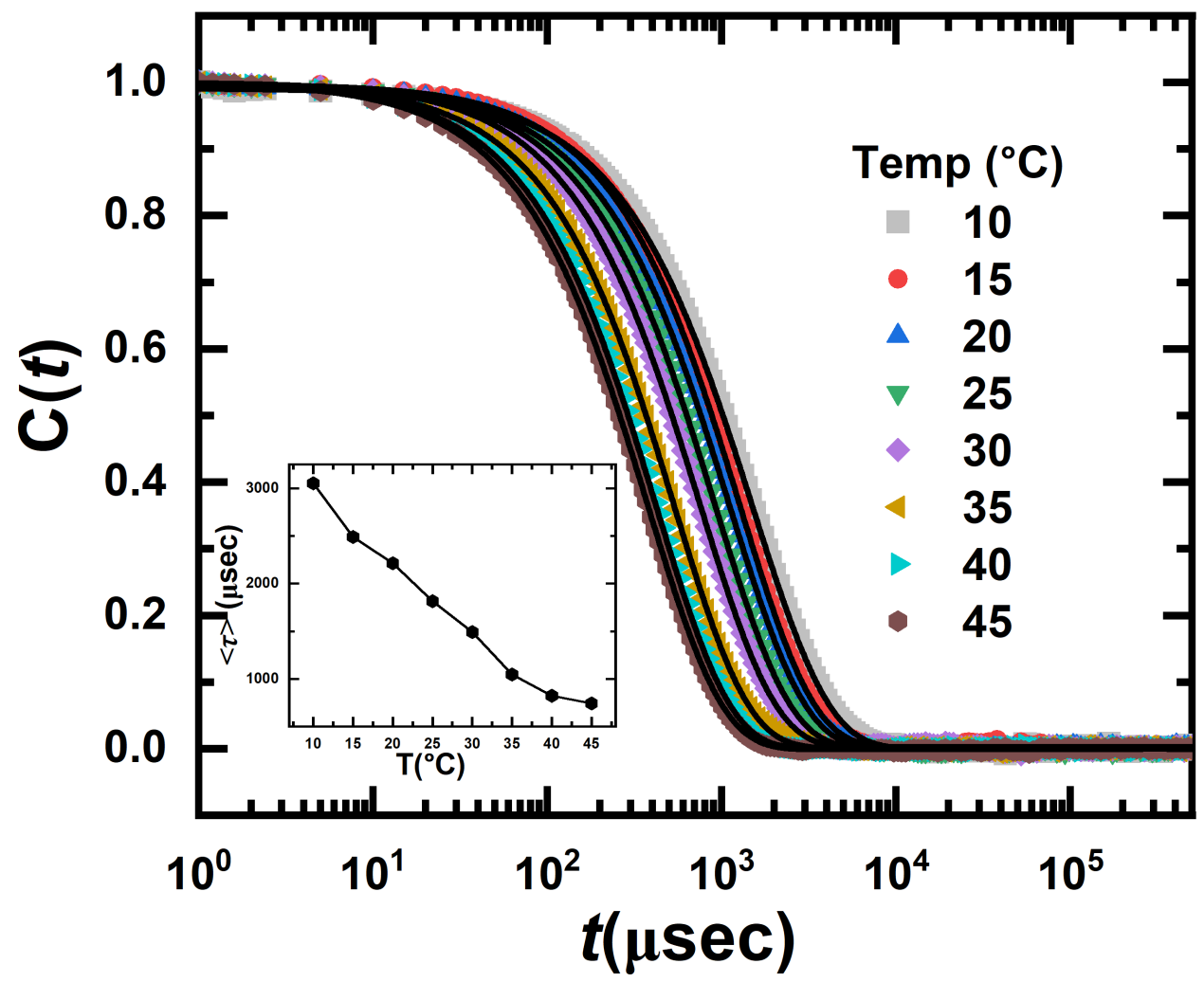


Fig. S3: **The normalized intensity autocorrelation functions, C($t$), as a function of delay time, $t$, for suspensions of PNIPAM particles (0.01% w/v) composed of 15% crosslinker-monomer concentration ratio. The black solid lines are fits to the expression: C($t$) = $[exp(-(\frac{t}{\tau})^{\beta})]^2$. The inset shows the variation of $<\tau>$ with temperature.**

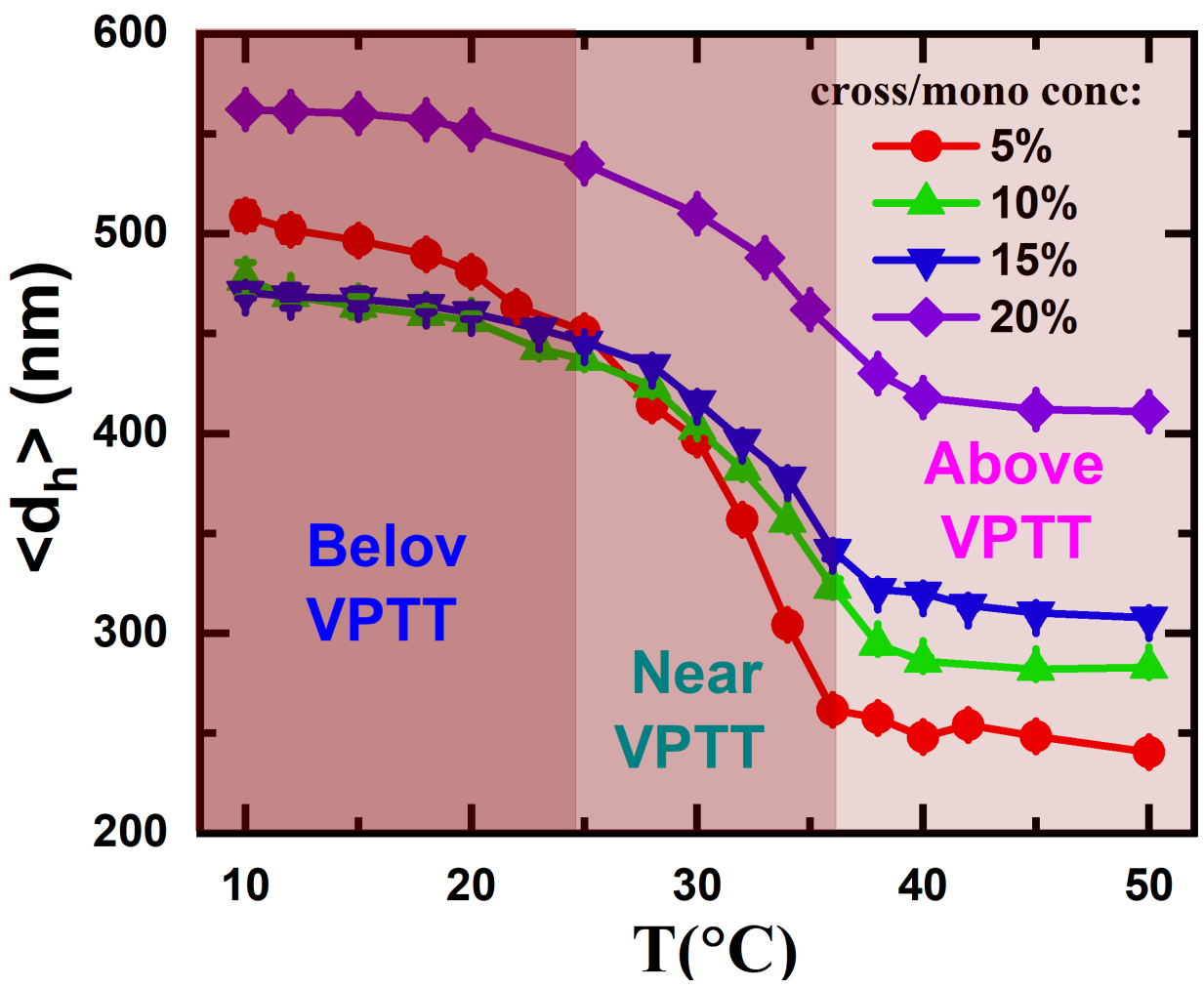


Fig. S4: **Temperature-dependent mean hydrodynamic diameters, <d$_\mathrm{h}$>, of PNIPAM particles of different stiffnesses.**

## ST3 Determination of the effective volume fraction, $\phi_{eff}$, of dense PNIPAM suspensions:

We estimated effective volume fractions, $\phi_{eff}$, of aqueous suspensions of PNIPAM particles prepared at pre-determined concentrations, c, ranging from 0.01 wt% to 0.9 wt%. A double gap geometry (DG-26.7) with a gap of 1.886 mm and an effective length of 40 mm was used to measure the viscosity, $\eta$, of these dilute suspensions as a function of strain rate, $\dot{\gamma_0}$, at a constant temperature 20°C. The data was fitted to the Cross model: $\frac{\eta-\eta_\infty}{\eta_0-\eta_\infty} = \frac{1}{1+(k\dot{\gamma_0})^m}$, where $\eta_0$ and $\eta_\infty$ correspond to the low and high shear rate viscosity plateaus respectively, $k$ is a time constant related to the relaxation time of the microgel in aqueous suspension, and $m$ is a dimensionless exponent [2]. The inset of Fig. S5 shows representative plots of the concentration dependence of $\eta$ as a function of $\dot{\gamma_0}$ for PNIPAM suspensions composed of particles with 15% crosslinker-monomer concentration ratio. The black lines are the fits of the data to the Cross model. The parameter $\eta_0$ was extracted from these fits, and the relative viscosities, $\eta_{rel}$, at various particle concentrations were calculated as $\eta_{rel} = \frac{\eta_0}{\eta_a}$. In this expression, $\eta_a$ is the viscosity of water at 20°C. Figure S5 shows a plot of $\eta_{rel}$ as a function of c for particles of different stiffnesses. The black lines are fits to Batchelor's equation: $\eta_{rel} = 1 + 2.5(\kappa\mathrm{c}) + 5.9(\kappa\mathrm{c})^2$ [3,4]. The parameter, $\kappa$, for particles of different stiffnesses was extracted from the fits to the $\eta_{rel}$ vs c data. We note that $\kappa$ is a shift factor that converts the mass concentration of PNIPAM to the effective volume fraction.

PNIPAM suspensions of desired effective volume fraction were prepared using the relation $\phi_{eff} =$

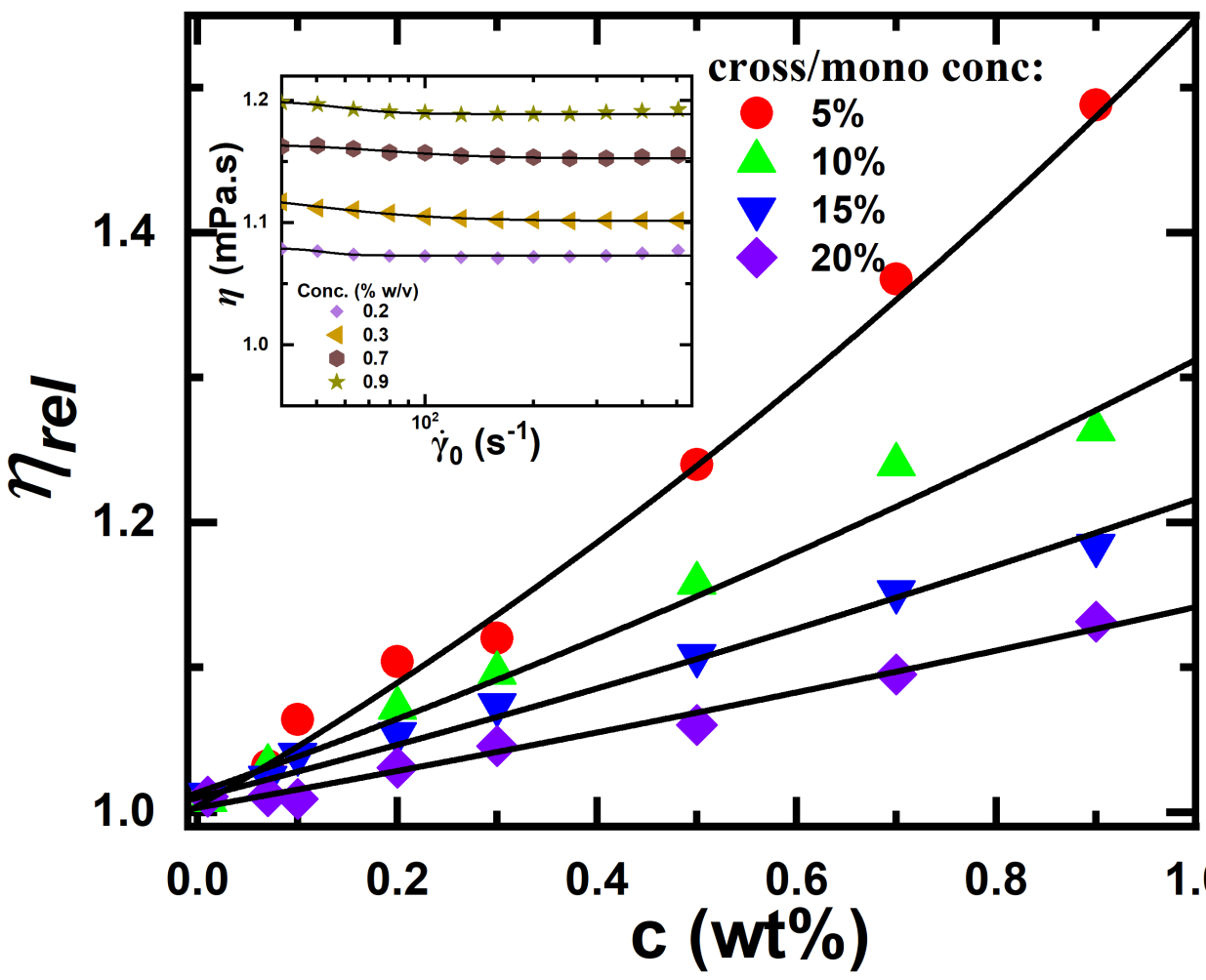


Fig. S5: **Relative viscosity, $\eta_{rel}$, as a function of particle concentration, c, of dilute suspensions of PNIPAM particles composed of different stiffnesses in water. The black lines are fit to the Batchelor's equation. The inset shows the variation of $\eta$ as a function of $\dot{\gamma_0}$ for aqueous suspensions of PNIPAM particles synthesized with 15% crosslinker-monomer concentration ratio, and the fits of the data to the Cross model.**

$\kappa c$. All experiments reported in this study were performed at two effective volume fractions, $\phi_{eff} = 1.4$ and 1.7 prepared at 20°C, using aqueous suspensions of PNIPAM particles with distinct stiffnesses. The temperature dependence of $\phi_{eff}$ was estimated using the equation $\phi_{eff}(\mathrm{T}) = \phi_{eff}(20°\mathrm{C}) \times \left[\frac{<\mathrm{d_h(T)}>}{<\mathrm{d_h(20°C)}>}\right]^3$ [5] and is shown in Fig. S6.

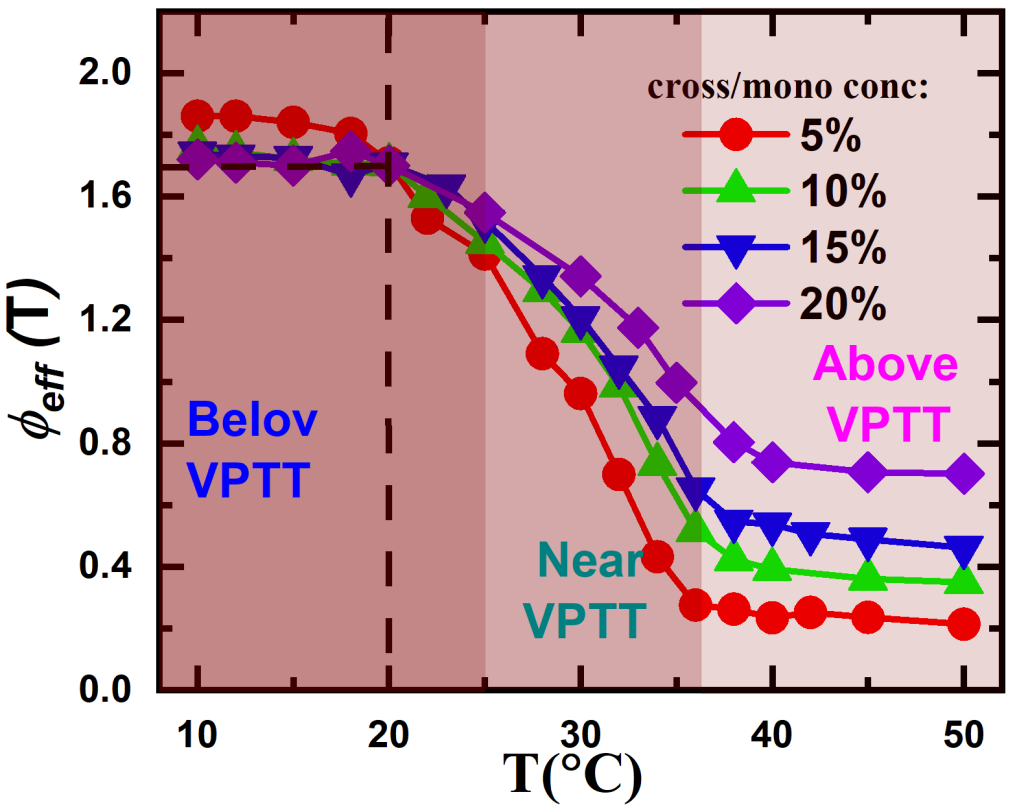


Fig. S6: **Temperature dependent effective volume fractions, $\phi_{eff}$, for dense aqueous PNIPAM suspensions composed of particles of different stiffnesses. The samples were prepared at $\phi_{eff} = 1.7$ at 20°C, highlighted by the vertical black dashed line.**

## ST4 Temperature-dependent dielectric relaxation behavior:

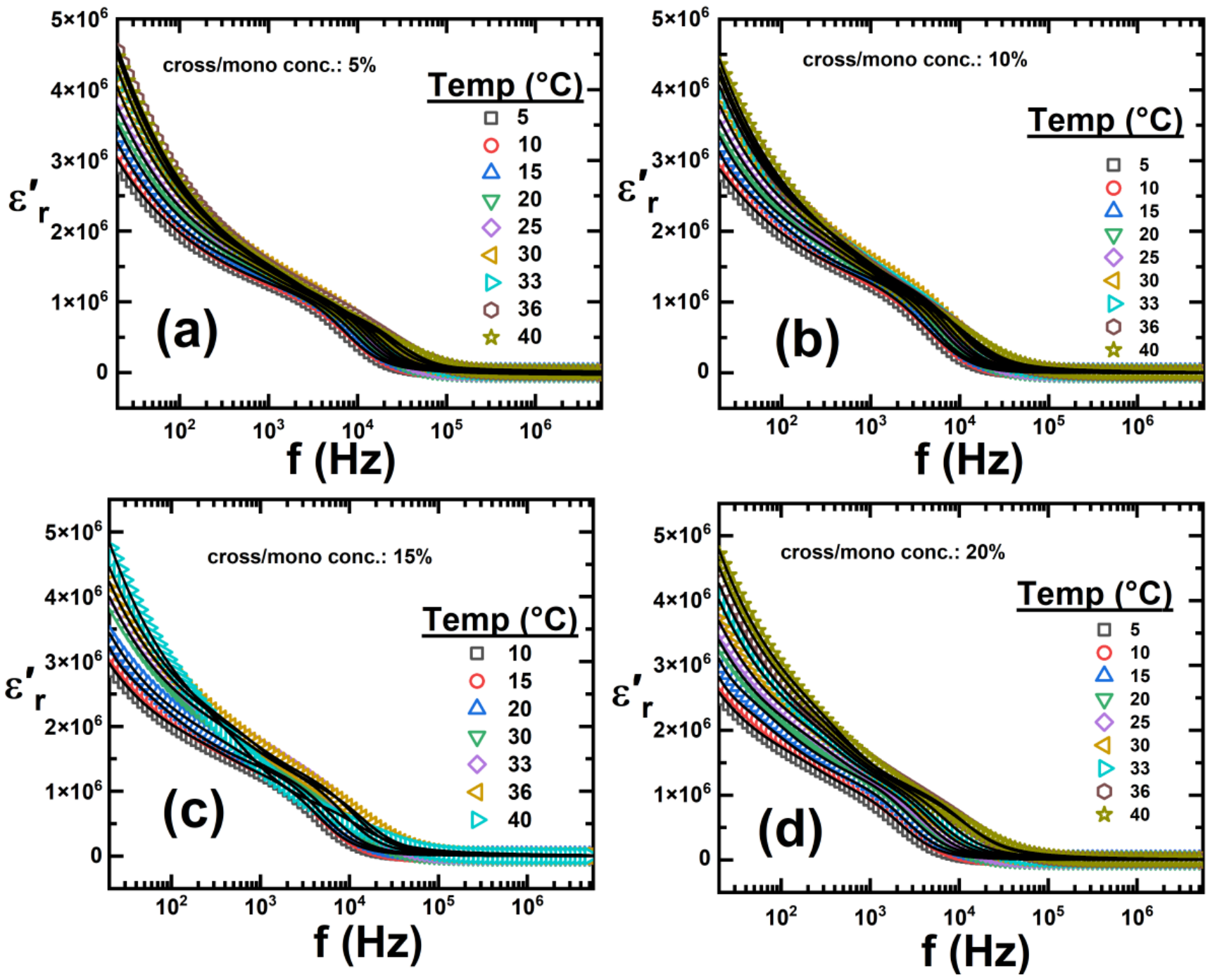


Fig. S7: **Frequency response of the real part of the relative dielectric permittivity, $\epsilon'_r$, of dense aqueous suspensions of PNIPAM particles ($\phi_{eff} = 1.7$ prepared at 20°C) of (a) 5%, (b) 10%, (c) 15% and (d) 20% crosslinker-monomer concentration ratios as a function of temperature. The black lines are fits to the real part of the Cole-Davidson model given in Eqn. 1 of the main paper.**

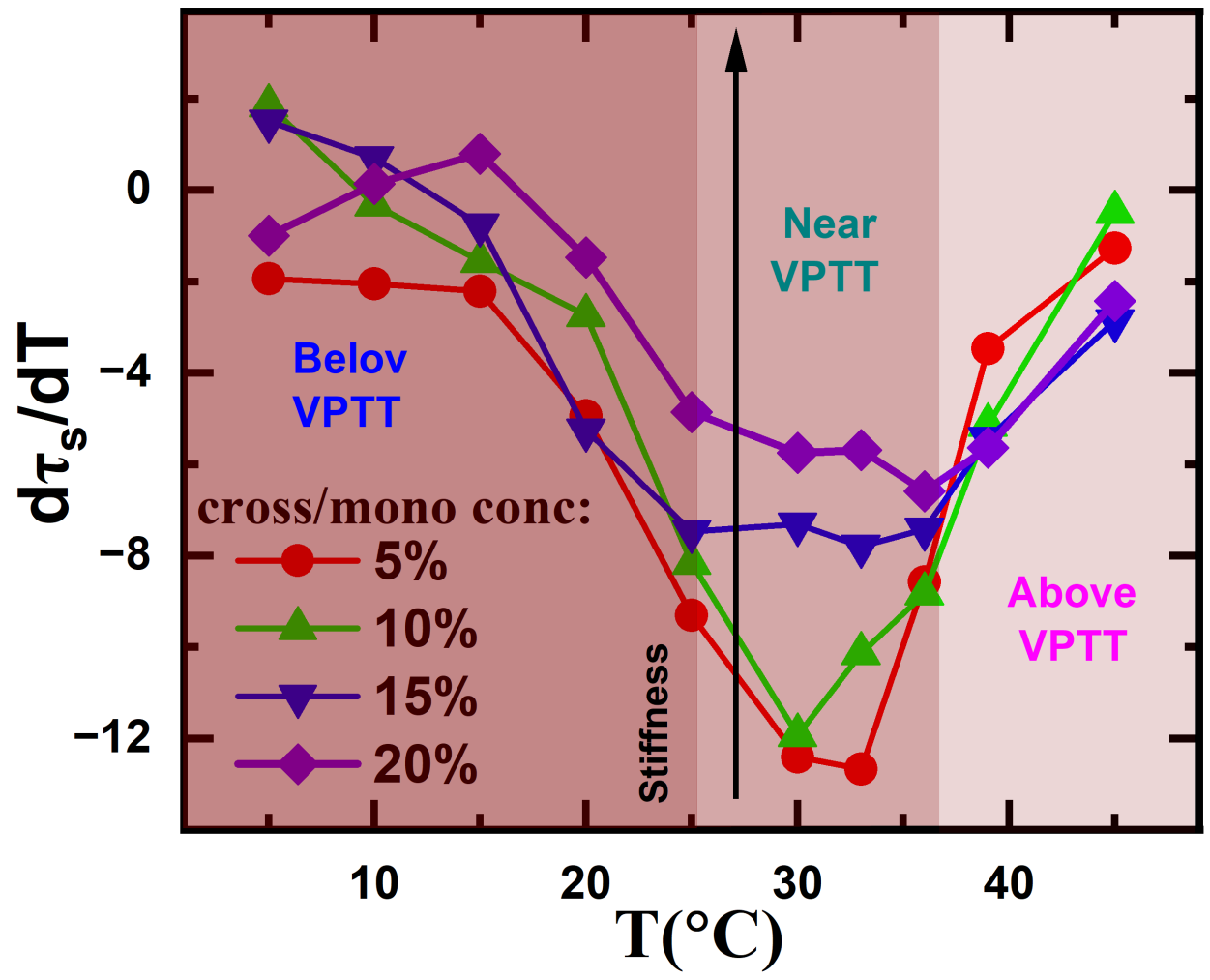


Fig. S8: **The temperature derivative of $\tau_s$ as a function of temperature of dense aqueous suspensions of PNIPAM particles ($\phi_{eff}$ = 1.7 prepared at 20°C) composed of different particle stiffnesses. Black arrow in the figure points toward increasing particle stiffness.**

# ST5 Strain-dependent dielectric relaxation behavior below the VPTT:

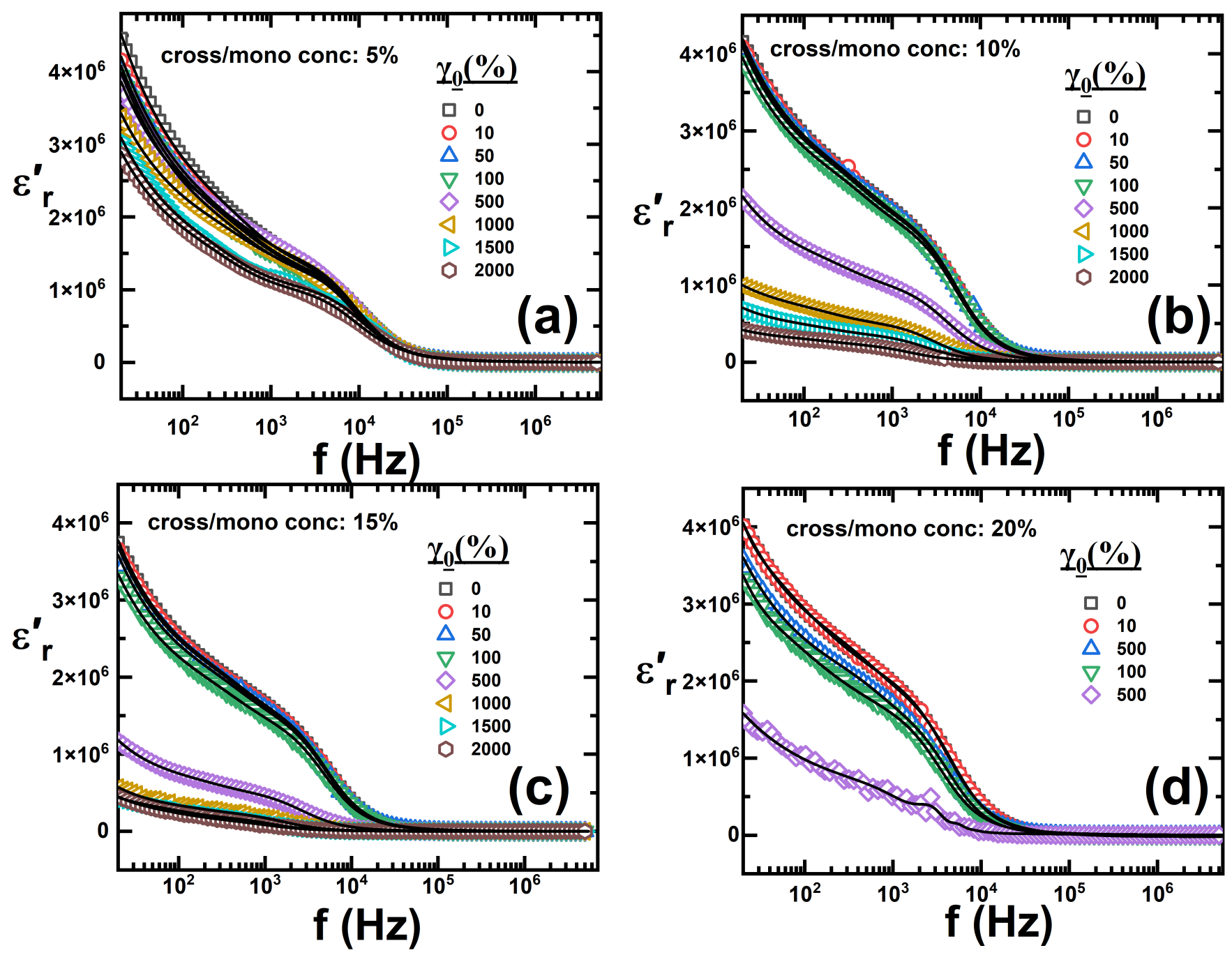

Fig. S9: (a) Frequency response of the real part of the relative dielectric permittivity, $\epsilon'_r$, of dense aqueous suspensions of PNIPAM particles ($\phi_{eff}$ = 1.7 prepared at 20°C) composed of (a) 5%, (b) 10%, (c) 15% and (d)20% crosslinker-monomer concentration ratios below the VPTT (20°C). The black lines are fits to the real part of the Cole-Davidson model given in Eqn. 1 of the main paper.

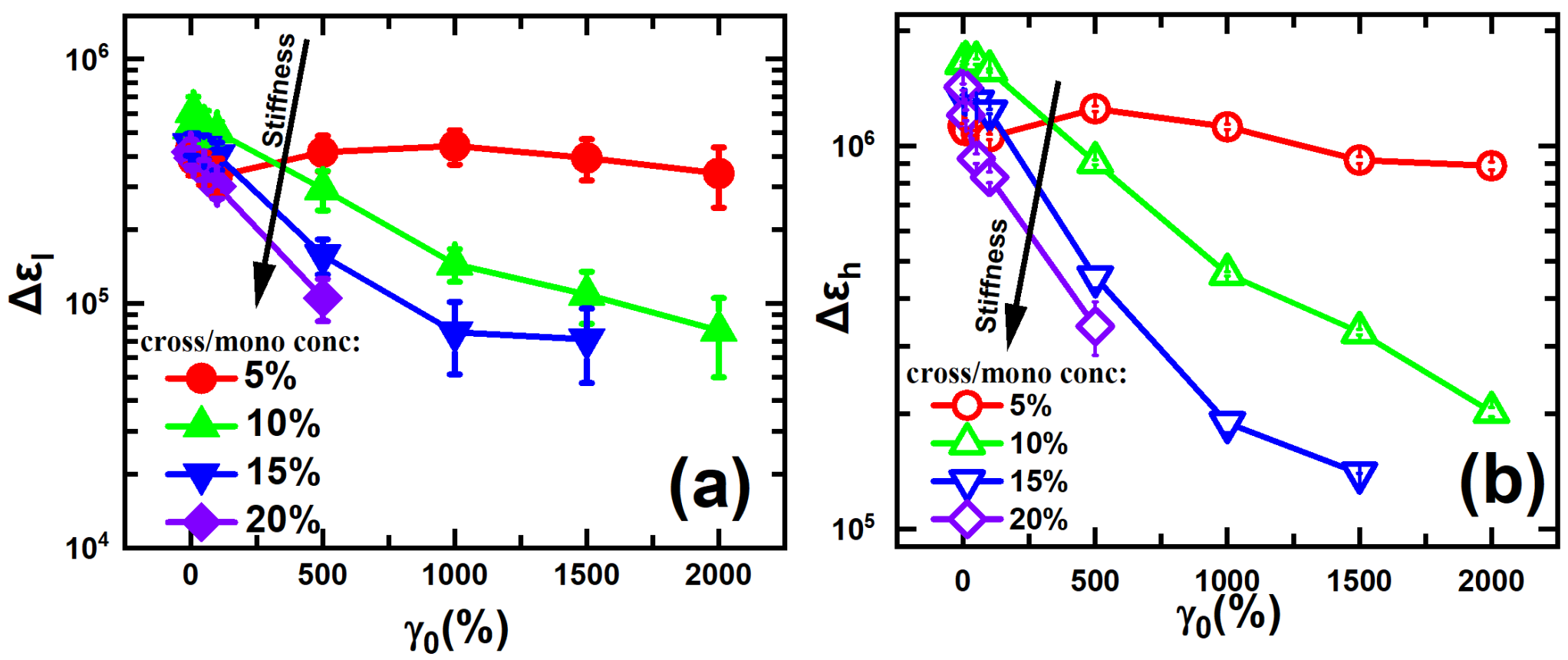

Fig. S10: (a) Low frequency, $\Delta\epsilon_l$, and (b) high frequency, $\Delta\epsilon_h$, dielectric relaxation strengths of dense aqueous suspensions of PNIPAM particles ($\phi_{eff}$ = 1.7 prepared at 20°C) composed of particles of different stiffnesses as a function of strain at 20°C. Black arrows point toward increasing particle stiffness.

# ST6 Strain-dependent dielectric relaxation behavior above the VPTT:

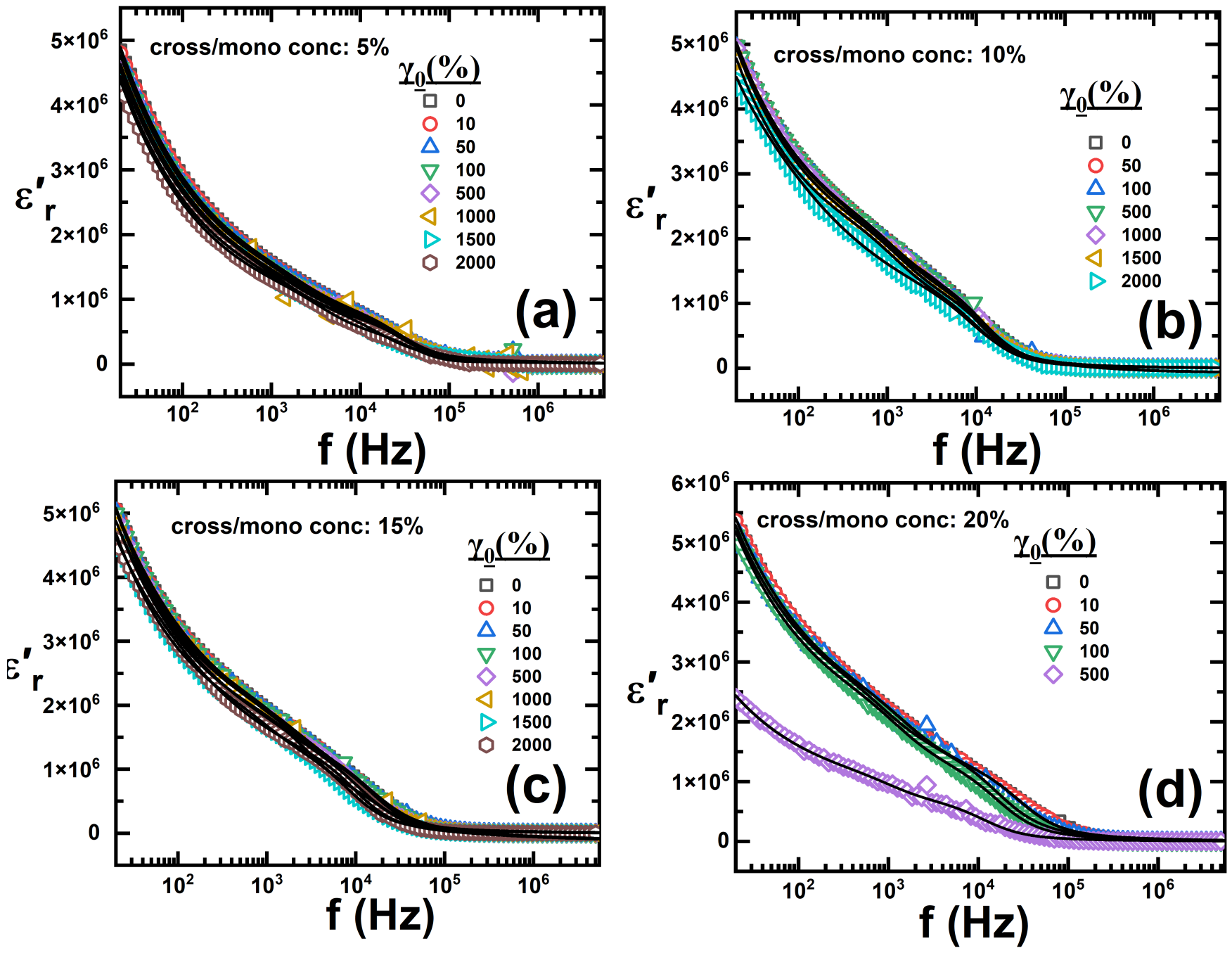


Fig. S11: (a) Frequency response of real part of relative dielectric permittivity, $\epsilon_r'$, of dense aqueous suspensions ($\phi_{eff} = 1.7$ prepared at 20°C) of PNIPAM particles composed of (a) 5%, (b) 10%, (c) 15% and (d) 20% crosslinker-monomer concentration ratios above the VPTT (45°C). The black lines are fits to the real part of the Cole-Davidson model given in Eqn. 1 of the main paper.

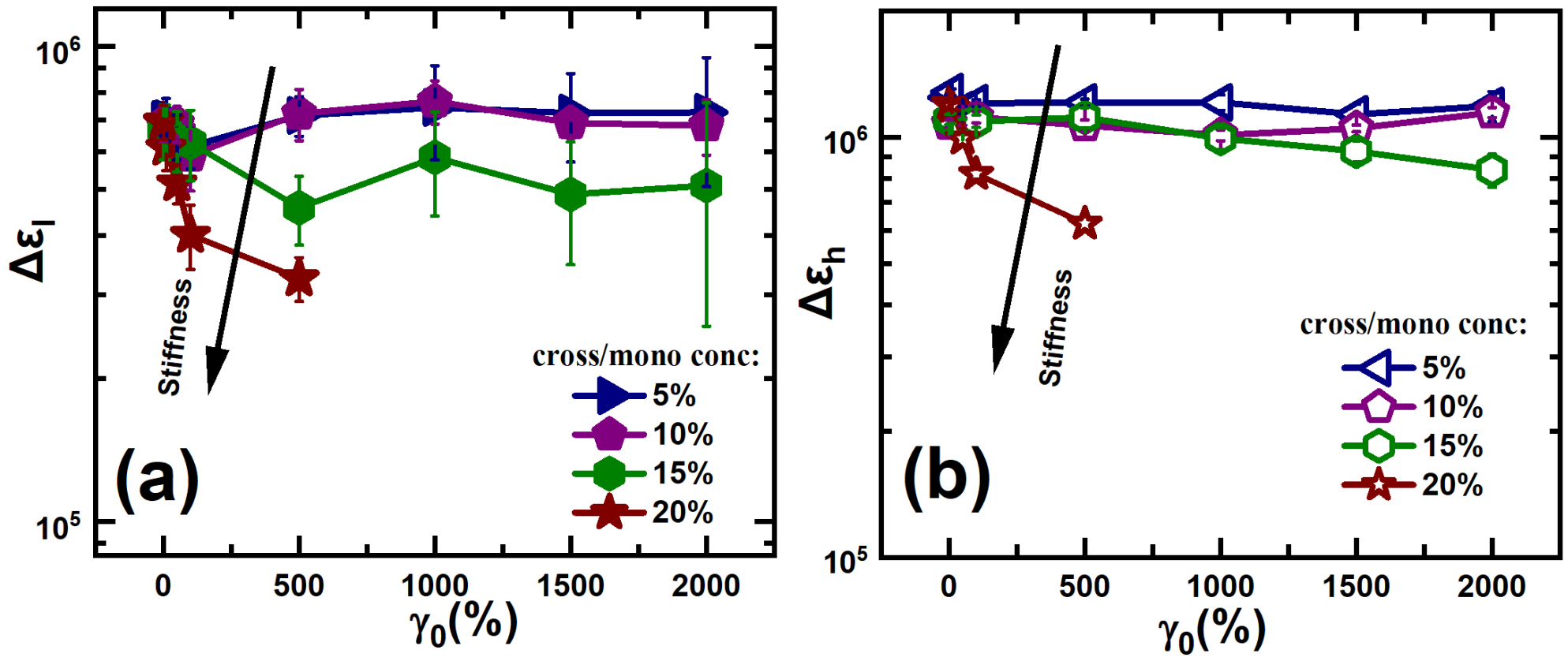


Fig. S12: (a) Low frequency, $\Delta\epsilon_l$, and (b) high frequency, $\Delta\epsilon_h$, dielectric relaxation strengths of dense aqueous suspensions ($\phi_{eff} = 1.7$ prepared at 20°C) of PNIPAM particles of different stiffnesses as a function of strain at 45°C. Black arrows point toward increasing particle stiffness.

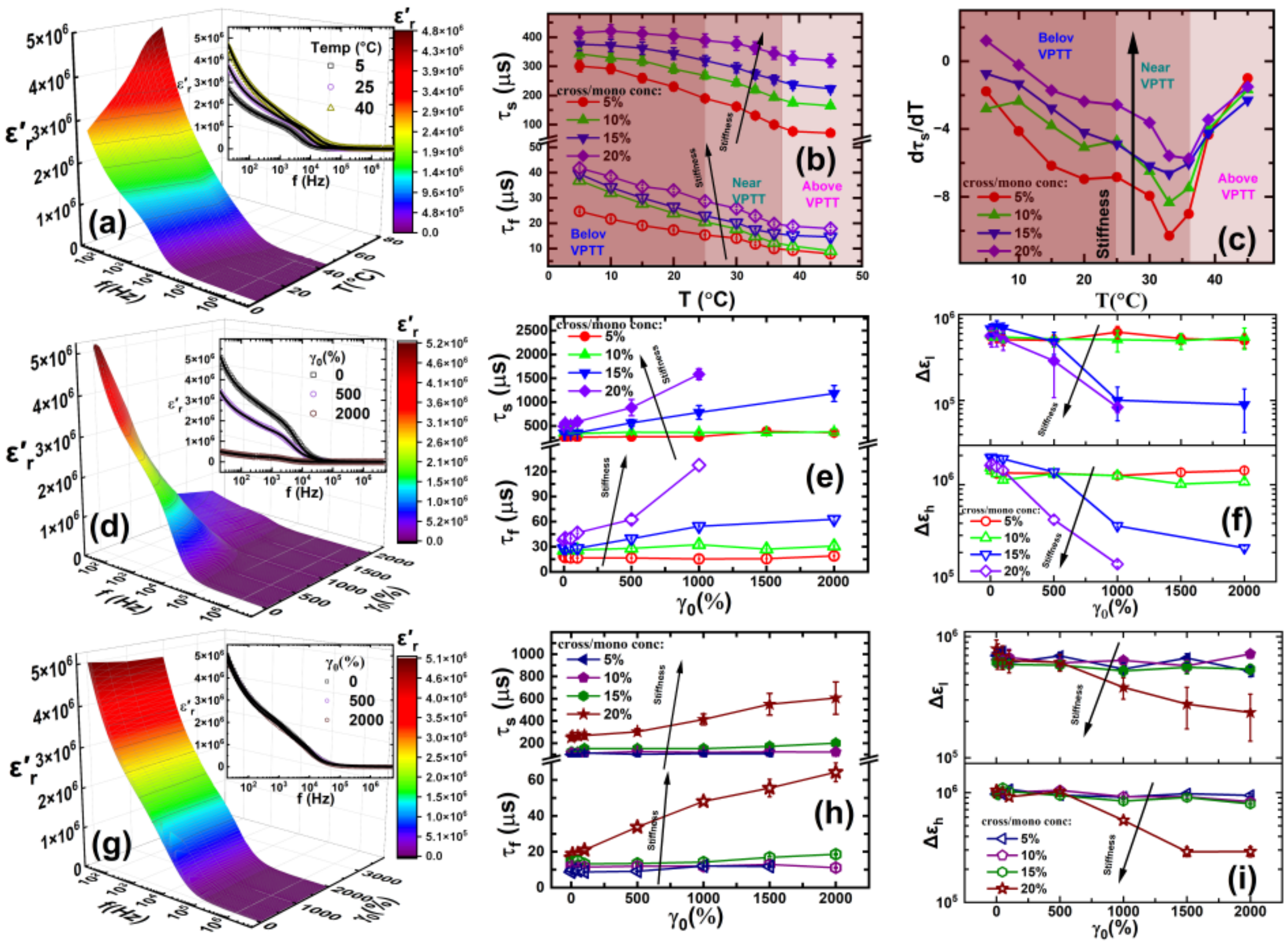


Fig. S13: (a) Three-dimensional representation of the real part of the relative dielectric permittivity, $\epsilon_r'$, of dense aqueous suspensions ($\phi_{eff} = 1.4$ prepared at 20°C) of PNIPAM particles, composed of 15% crosslinker-monomer concentration ratio, as a function of applied frequency and temperature under zero-strain condition. The inset shows $\epsilon_r'$ vs frequency at three different temperatures and the black lines are fits to the real part of the Cole-Davidson model given in Eqn. 1 of the main paper, (b) temperature-dependent slow relaxation timescales, $\tau_s$ (bold symbols), and fast relaxation timescales, $\tau_f$ (hollow symbols), and (c) temperature-dependent $d\tau_s/dT$. (d) Three-dimensional representation of the real part of the relative dielectric permittivity, $\epsilon_r'$ as a function of applied frequency and strain below the VPTT (20°C) of the same sample as used in (a). The inset shows $\epsilon_r'$ vs frequency at three different strains and the black lines are fits to the real part of the Cole-Davidson model, (e) strain-dependent $\tau_s$ and $\tau_f$ below the VPTT (20°C) and (f) strain-dependent low, $\Delta\epsilon_l$, and high $\Delta\epsilon_h$ frequency dielectric strengths below the VPTT (20°C). The corresponding strain-dependent data above the VPTT (45°C) is shown in (g), (h) and (i) respectively. Black arrows in figures (b), (c), (e), (f), (h) and (i) point toward increasing particle stiffness.

## ST7 Cryogenic field emission scanning electron microscopy (cryo-FESEM)

Cryo-FESEM imaging was performed using a field emission scanning electron microscope (Ultra Plus FESEM-4098, Carl Zeiss) operated at an electron beam strength of 5 KeV. The samples were first equilibrated at 45°C, mounted on a rivet attached to a cryostub, and rapidly vitrified in liquid nitrogen slush at -190°C. They were then transferred to a cryo-preparation chamber (PP3000T, Quorum Technologies) maintained at -190°C. The vitrified samples were sublimated at -90°C for 5 min, cryo-fractured using the chamber's built-in stainless steel knife, and further sublimated for 12 min at -90°C to remove surface ice. Subsequently, a platinum coating of thickness around 5 nm was applied to the sample surfaces to ensure good image contrast. The samples were then transferred to the microscope cryo-stage maintained at -190°C for imaging. Surface images were acquired using back-scattered secondary electrons.

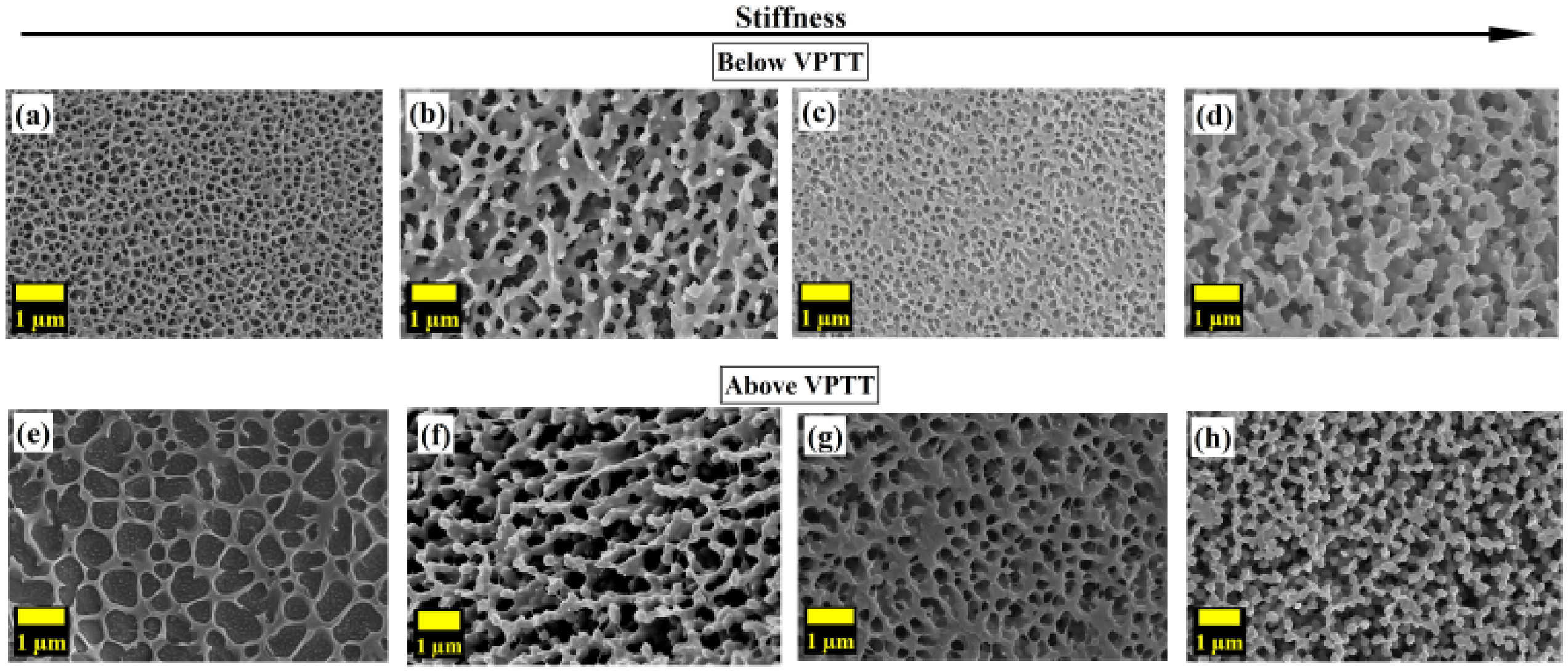


Fig. S14: **Cryo-FESEM micrographs of dense aqueous suspensions of PNIPAM microgel particles ($\phi_{eff}$ = 1.7 prepared at 20°C) of increasing stiffness having crosslinker-monomer concentration ratios: (a) 5%, (b) 10%, (c) 15%, and (d) 20% below the VPTT. The images of the same samples above the VPTT are shown in (e)-(h), respectively.**

## ST8 Bulk stress relaxation behavior below the VPTT:

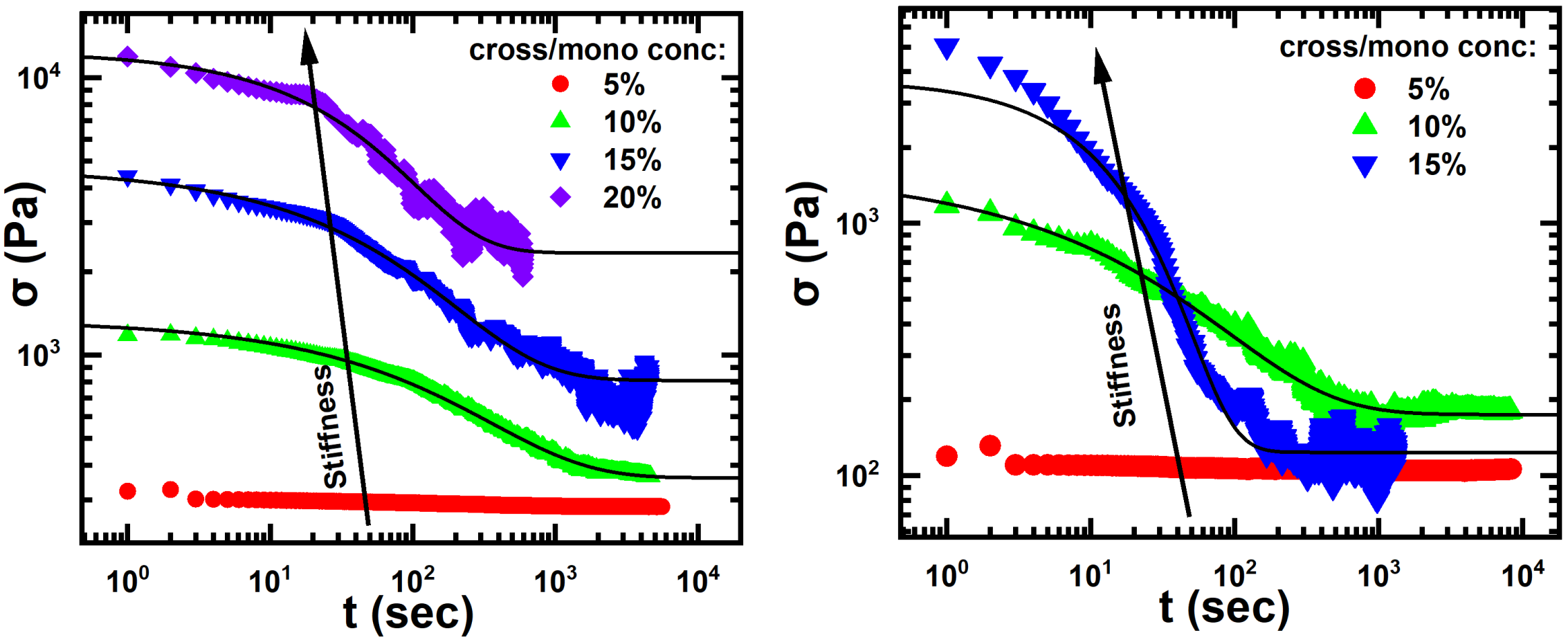


Fig. S15: **Bulk stress relaxation versus time below the VPTT for dense PNIPAM suspensions ($\phi_{eff}$ = 1.7 prepared at 20°C) constituted by particles of distinct stiffnesses, when subjected to different large oscillatory strain of peak-to-peak amplitudes, $\gamma_0$: (a) 300% and (b) 1000%. Black arrows point toward increasing particle stiffness.**

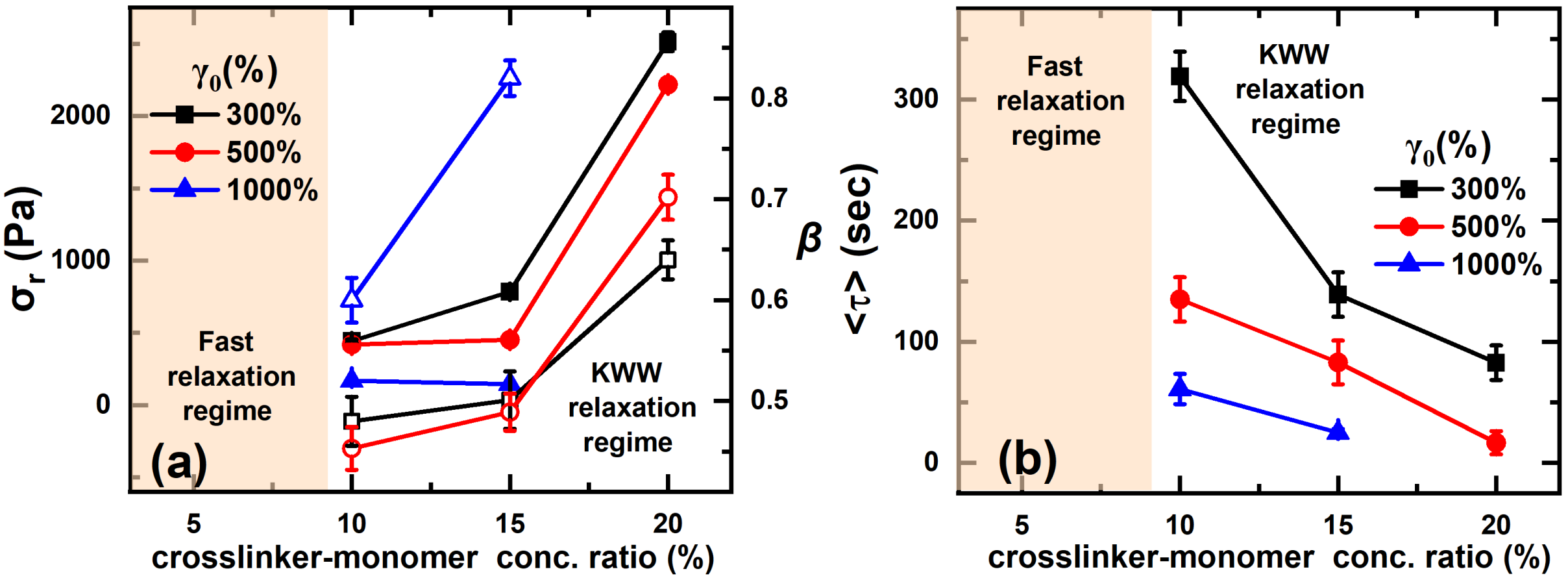


Fig. S16: **(a) Variation of residual stress, $\sigma_r$ (bold symbols), and stretching exponents, $\beta$ (hollow symbols), as a function of particle stiffness for different large oscillatory strain amplitudes in dense suspension ( $\phi_{eff}$ = 1.7 prepared at 20°C). (b) Bulk stress relaxation timescales, $< \tau >$, as a function of particle stiffness at 20°C for different large oscillatory strain amplitudes in the same suspensions.**

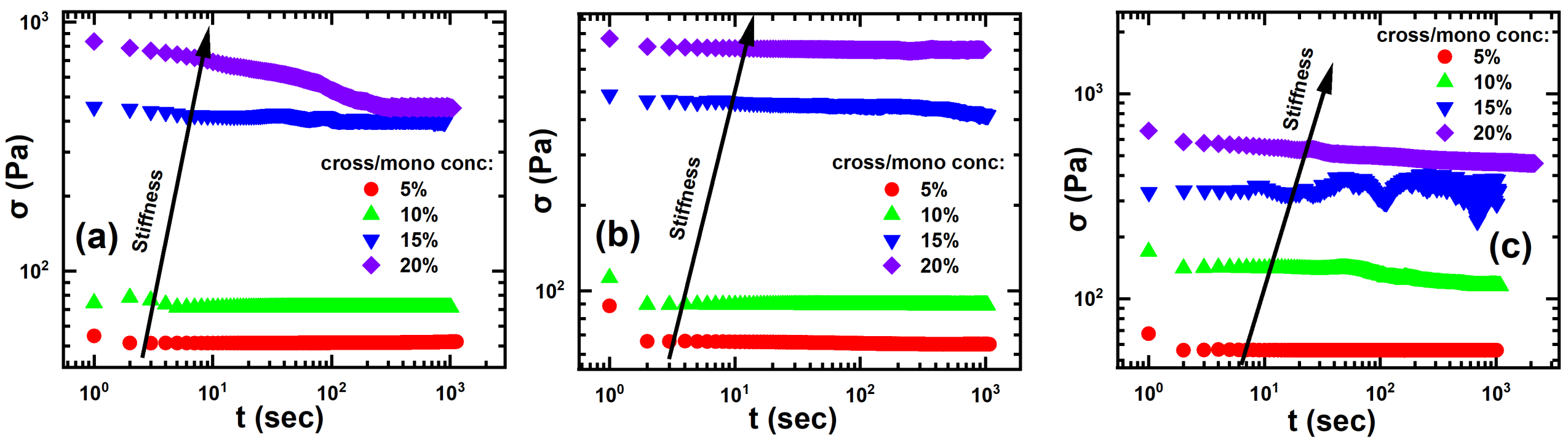


Fig. S17: **Bulk stress relaxation versus time above the VPTT for dense PNIPAM suspensions constituted by particles of different stiffnesses ($\phi_{eff}$ = 1.4 prepared at 20°C), when subjected to different $\gamma_0$: (a) 300%, (b) 500% and (c) 1000%. Black arrows point toward increasing particle stiffness.**

## ST9 Bulk stress relaxation behavior above the VPTT:

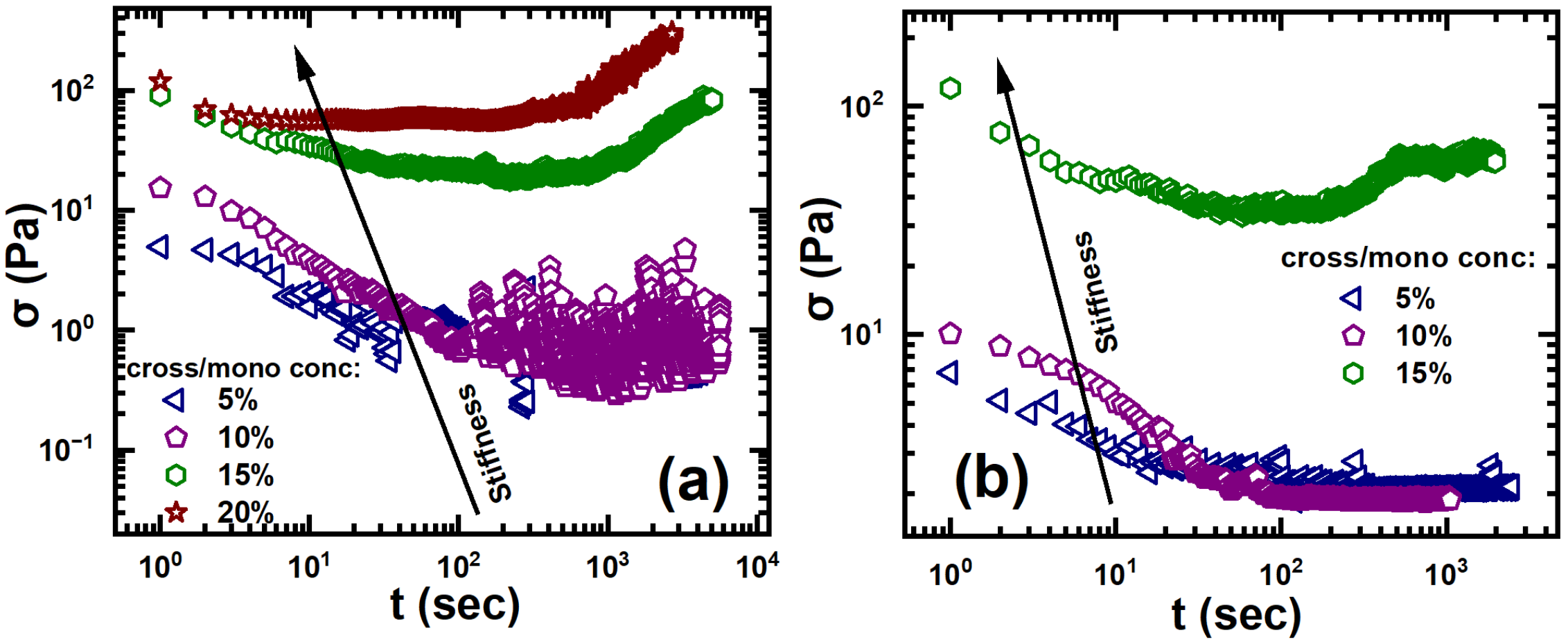


Fig. S18: **Bulk stress relaxation versus time above the VPTT for PNIPAM suspensions ($\phi_{eff}$ = 1.7 prepared at 20°C) constituted by particles of different stiffnesses, when subjected to different $\gamma_0$: (a) 300% and (b) 1000%. Black arrows point toward increasing particle stiffness.**

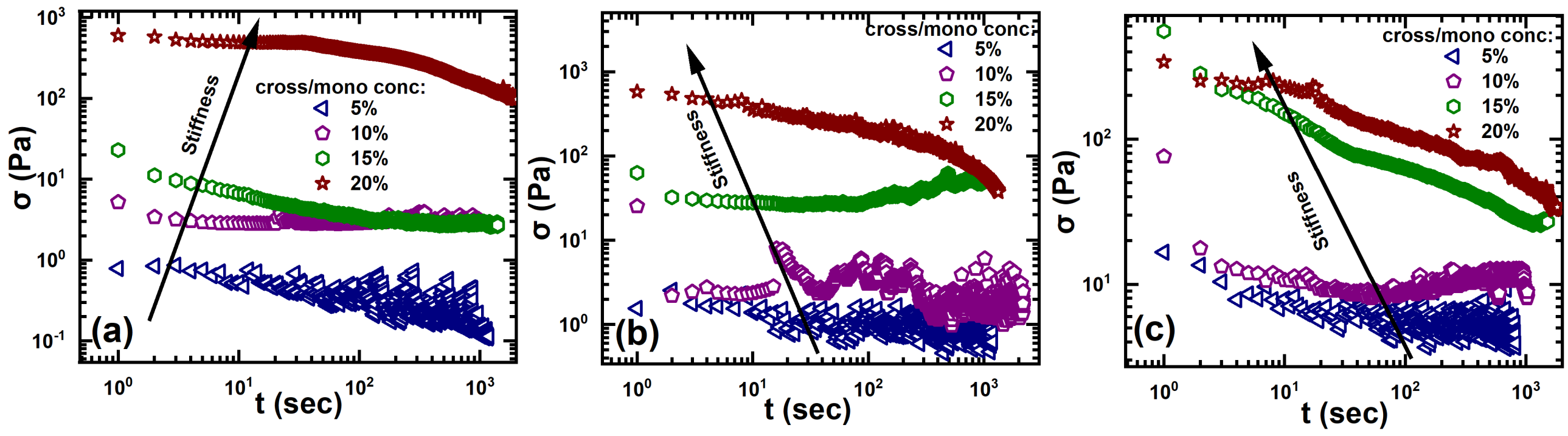


Fig. S19: **Bulk stress relaxation versus time above the VPTT for dense PNIPAM suspensions constituted by particles of different stiffnesses ($\phi_{eff}$ = 1.4 prepared at 20°C), when subjected to different $\gamma_0$: (a) 300%, (b) 500% and (c) 1000%. Black arrows point toward increasing particle stiffness.**


\end{document}